\documentclass[final,5p,times,twocolumn]{elsarticle}
\usepackage{amssymb}
\usepackage{amsmath}
\usepackage{graphicx}
\usepackage{mathtools}
\usepackage[super]{cite}
\usepackage{xcolor}
\usepackage[verbose,hypertexnames=false]{hyperref}
\usepackage{subcaption}
\usepackage{xfrac}
\usepackage{cancel}
\usepackage{xspace}

\newcommand{\loopcalculus}{smooth--variation loop calculus\xspace}

\newcommand{\Mathematica}{\textit{Mathematica\textsuperscript{\resizebox{!}{0.8ex}{\textregistered}}}}
\def\8{\infty}
\def\oh{\sfrac{1}{2}}

\newcommand*{\I}{\imath}%

\def\eps{\epsilon}

\def\const{\textit{ const }}
\def\undertext#1{\vtop{\hbox{#1}\kern 1pt \hrule}}
\def\Ra{\Rightarrow}

\def\abs#1{\left| #1\right|}

\def\pd#1{\partial_{#1}}
\def\VEV#1{\left\langle #1\right\rangle}

\def\tr{\hbox{tr}\,}

\def\pbyp#1#2{\frac{\partial#1}{\partial#2}}
\def\ff#1{\frac{\delta}{\delta#1}}
\def\fbyf#1#2{\frac{\delta#1}{\delta#2}}

\def\bea{\begin{eqnarray} && &&}
\def\eea{\end{eqnarray}}

\def\rf#1{(\ref{#1})}
\let\oldexp\exp
\renewcommand{\exp}[1]{\oldexp\left(#1\right)}

\def\NS{Navier-Stokes}

\def\ral{r_{\alpha}}

\newcommand{\Mod}[1]{\ (\mathrm{mod}\ #1)}

\def\XXint#1#2#3{{\setbox0=\hbox{$#1{#2#3}{\int}$}
     \vcenter{\hbox{$#2#3$}}\kern-.5\wd0}}

\renewcommand{\Re}{\textbf{Re }}
\renewcommand{\Im}{\textbf{Im }}

\newcommand{\tpmod}[1]{{\@displayfalse\Mod{#1}}}
\usepackage[english]{babel}

\usepackage{mathtools}

\DeclarePairedDelimiter\floor{\lfloor}{\rfloor}
\newcommand{\pctWPDF}[3]{
\begin{figure}
    \centering
    \includegraphics[width=#1\textwidth]{#2.pdf}
    \caption{#3}
    \label{fig::#2}
\end{figure}
}

\def\lb#1{\left[#1\right.}
\def\rb#1{\left.#1\right]}
\DeclarePairedDelimiterX{\infdivx}[2]{(}{)}{%
  #1\;\delimsize\|\;#2%
}

\DeclarePairedDelimiter{\norm}{\lVert}{\rVert}

\def\YM{Yang--Mills}
\usepackage{etoolbox} % for patching environments if needed
\usepackage{amsthm}
\makeatletter
\newenvironment{smalleq}{%
  \begingroup
  \small
  \addtolength{\jot}{.2em}%
  \setlength{\arraycolsep}{1.2pt}%
  \begin{eqnarray}%
}{%
  \end{eqnarray}%
  \endgroup
}
\makeatother

\makeatletter

\def\lb#1{\left[#1\right.}
\def\rb#1{\left.#1\right]}
\newtheorem{theorem}{Theorem}
\usepackage{etoolbox} % for patching environments if needed

\journal{Nuclear Physics B}

\begin{document}

\begin{frontmatter}

\title{Spontaneous quantization of the Yang--Mills gradient flow}

\author[aff1]{Alexander Migdal\corref{cor1}}
\ead{amigdal@ias.edu}
\ead[url]{https://orcid.org/0000-0003-2987-0897}

\cortext[cor1]{Corresponding author}
\affiliation[aff1]{organization={Institute for Advanced Study}, city={Princeton}, country={USA}}

\begin{abstract}
We formulate a nonsingular loop--space calculus for the Yang--Mills (YM) gradient flow
\emph{directly in terms of Wilson loop functionals}, rather than the underlying gauge fields.
All variations act within the manifold of smooth loops via ``dot derivatives''
that are finite, parametrization--invariant, and free of cusp or backtracking singularities.
This yields a closed \emph{linear} diffusion equation in loop space for Wilson loops.
The associated loop operator is universal and trilinear in functional derivatives, and the formulation
automatically factors out gauge transformations, exposing the gauge--invariant core of the flow.
The construction is valid for any (Abelian or non--Abelian) gauge group.

We identify two distinct classes of exact solutions.
First, a self--dual (Hodge--dual) matrix--valued minimal surface whose area functional,
when exponentiated, solves the fixed--point loop equation exactly, without contact terms or ambiguities;
for planar loops the dual area equals $2\sqrt{2}$ times the Euclidean minimal area,
providing a geometrically grounded confinement mechanism.
We also prove that the ordinary minimal surface in $\mathbb{R}^4$ fails to satisfy the fixed--point loop equation,
due to a singular nonvanishing contribution from the loop operator.

Second, a decaying--flow solution in which the momentum loop executes a periodic random walk
on regular star polygons (the ``Euler ensemble'' known from Navier--Stokes turbulence).
\emph{Both} classes realize spontaneous quantization:
the self--dual solution furnishes a stationary quantized state (a fixed manifold of the flow),
while the decaying solution describes a quantized trajectory that approaches a pure--gauge vacuum along a universal path.

Thus we obtain exact solutions of the Wilson--loop evolution in YM gradient flow.
We discuss the emergence of quantum--like Wilson--loop statistics from deterministic classical dynamics,
potential implications for confinement in QCD, and the role of these fixed manifolds and trajectories as attractors
in the space of YM gradient--flow solutions.
\end{abstract}

\begin{keyword}
Gradient Flow \sep Yang--Mills theory\sep Loop equation \sep Number Theory\sep Euler ensemble
\PACS 03.65.$-$w, 04.62.+v
\end{keyword}

\end{frontmatter}
\hypertarget{sec-introduction-gradient-flow-in-non-abelian-gauge-theories}{}
\section{Introduction: Gradient Flow in Non-Abelian Gauge Theories}
\pdfbookmark[1]{Introduction: Gradient Flow in Non-Abelian Gauge Theories}{gradFlow}

The gradient flow in non-Abelian gauge theories provides a powerful framework for systematically studying the dynamics of gauge fields via an auxiliary time dimension. Originally introduced to enhance theoretical understanding and numerical stability in lattice gauge theories, gradient flows have since become an essential tool for exploring nonperturbative phenomena in quantum field theory.

The gradient flow for a non-Abelian gauge field \(A_\nu\) is governed by the equation
\begin{smalleq}\hypertarget{gradFlow}{}
\label{gradFlow}
\partial_\tau A_\nu = \alpha [D_\mu, F_{\mu \nu}],
\end{smalleq}
where \(D_\mu\) denotes the covariant derivative, \(F_{\mu \nu} = [D_\mu, D_\nu]\) is the non-Abelian field strength tensor, and \(\alpha\) is a positive flow parameter. This equation can be interpreted as a form of non-Abelian diffusion, with $\alpha$ acting as the diffusion coefficient.

This dissipative evolution monotonically decreases the Yang–Mills action, defined by
\begin{smalleq}\hypertarget{Energy}{}
\label{Energy}
E[A] = -\frac{1}{2} \int d^4x\, \text{tr}(F_{\mu \nu} F^{\mu \nu}).
\end{smalleq}

We consider the anti-Hermitian gauge field \(A_\mu\), where \(\mu = 1,\dots, 4\), as valued in the Lie algebra of a semi-simple gauge group \(\mathcal{G}\). Although this field may be expanded in terms of algebra generators, we will not require an explicit expansion here.

In recent decades, gradient flows—especially those in Yang–Mills theory—have attracted significant attention in both mathematics and theoretical physics. These flows are closely related to the stochastic quantization of Yang–Mills theory, a cornerstone of modern particle physics. Simultaneously, they provide profound insights into the topology of gauge field spaces, thereby contributing to advancements in contemporary geometry and topology. For a recent review, see the monograph~\cite{Feehan2016}.

It is well known that self-dual and anti-self-dual gauge configurations (instantons and anti-instantons) are fixed points of the Yang–Mills flow in four dimensions. Moreover, it has been proven~\cite{Waldron2019} that no finite-time singularities can arise in 4D Yang–Mills gradient flows.

In contrast, for the analogous problem of incompressible \NS{} flow in three dimensions, the existence of finite-time singularities remains unresolved. The leading nontrivial alternative appears to be spontaneous stochasticity—the emergence of a statistical distribution of the field from a deterministic partial differential equation. This phenomenon lies at the heart of turbulence, the well-known and ubiquitous behavior observed across natural systems.

Recently, we have made significant progress \cite{migdal2023exact, migdal2024quantum} in understanding turbulence by identifying a class of spontaneously stochastic solutions to the Navier–Stokes equations. This class, distinguished by a discrete number-theoretic structure, is what we call the Euler ensemble. Remarkably, the continuum limit of the Euler ensemble is a solvable string theory with a discrete target space. 

The resulting energy spectrum and other observables of decaying turbulence were analytically computed in quadrature\cite{migdal2024quantum}. We computed the whole spectrum of universal critical indexes -- the powers of the decay of correlation functions and the energy spectrum. Some of these indexes are rational numbers, others come in complex-conjugate pairs, related to the nontrivial zeros of the Riemann zeta function.

In this paper, we discover spontaneous stochasticity in the non-Abelian gradient flow, equivalent to spontaneous quantization of a classical gauge theory. We identify emergent turbulent dynamics that give rise to quantum structures. Specifically, Wilson loops—whether representing classical configurations or those sampled from a Yang–Mills distribution—evolve under the flow toward configurations that can be interpreted as string theories with compact target spaces.

We find two distinct nonperturbative solutions to the QCD gradient flow: one exhibiting turbulent decay and another corresponding to a topological fixed point. These results suggest a fundamental mechanism by which quantum randomness and structure may emerge spontaneously from deterministic, nonlinear dynamics. Furthermore, the analytic nature of these solutions, which go beyond perturbation theory, opens new avenues for understanding quark confinement.

The remainder of this paper is devoted to presenting and exploring these results.

In parallel with this article, we maintain a ``living guide'' that presents the main structures of the theory in a hierarchical and accessible format. 
This guide is hosted at \url{https://sashamigdal.github.io/QuantumSolution/YangMillsGuide} 
and is intended to be browsed alongside the present paper---or, for a quick overview, on its own. 
It provides an interactive map of the key equations, exact solutions, and conceptual framework, 
and will be periodically updated to reflect future developments.

\hypertarget{sec-gradient-flow-of-the-wilson-loop}{}
\section{Gradient flow of the Wilson loop}
\pdfbookmark[1]{Gradient flow of the Wilson loop}{WCdef}

\hypertarget{sub-the-wilson-loop-as-generating-functional-for-field-strength}{}
\
\subsection{The Wilson loop as generating functional for field strength}
Our main variable will be the Wilson loop
\begin{smalleq}\hypertarget{WCdef}{}
\label{WCdef}
    W[C] = \VEV{\tr \hat P \exp{ \oint d \theta\, \dot C_\mu(\theta)\, A_\mu( C(\theta))}}_A,
\end{smalleq}
with averaging over the distribution of the gauge field. In QCD this averaging uses the YM functional measure
\begin{smalleq}
    \VEV{X[A]}_A = \frac{\int [\delta A]\, X[A]\,\exp{- \beta E[A]}}
                        {\int [\delta A]\,\exp{- \beta E[A]}},
\end{smalleq}
which can be interpreted as the Gibbs distribution of a statistical system in four dimensions, with inverse temperature $\beta = 1/T$. In the \YM{} theory, the temperature is equivalent to the bare coupling constant $T = g_0^2$. The regularization and gauge fixing necessary to define these functional integrals have been discussed extensively in the literature. 

Perturbatively (in $g_0$), dimensional regularization works for arbitrary flow time as well as for \YM{} theory at $\tau=0$. Beyond perturbation theory, lattice gauge theory provides a nonperturbative regularization. As we shall see below, the loop equation can be regularized in a different way, without breaking space symmetries (as the lattice does) and without gauge fixing (as dimensional regularization does).

Thus the Wilson loop is a gauge--invariant functional of $C \in \mathbb{R}^4$, serving both as a generating functional for expectation values of gauge field strength parallel--transported to the same point in space and as a directly observable amplitude for quarks interacting with the gauge field.  

\subsection{Gauge-invariant observables generated by Wilson loops}

Local, gauge-invariant operators are generated from the Wilson loop by choosing
appropriate loop shapes and taking area derivatives. The simplest case is the
\emph{shrinking} limit, in which the loop contracts to a point:
\begin{smalleq}
    &&W[C] \to N_c + \frac{ \VEV{\tr\left(\Sigma_{\mu\nu}(C) F_{\mu\nu}\right)^2}}{2 };\\
    && \Sigma_{\mu\nu}(C) = \oh \oint_C r_\mu \, d r_\nu\,,
\end{smalleq}
so that $\Sigma_{\mu\nu}(C)$ plays the role of the infinitesimal area bivector.

More informative are \emph{two-point} correlators. Taking two area derivatives at
separated points on the loop produces the gauge-invariant field-strength
correlation, with parallel transport along the loop between the insertion points:
\begin{smalleq}
    &&\VEV{\tr\!\Big[F_{\mu\nu}\!\big( C(\theta_1)\big)\,U_{\theta_1,\theta_2}\,
         F_{\alpha\beta}\!\big(C(\theta_2)\big)\,U_{\theta_2,\theta_1+ 2\pi}\Big]}
    = \nonumber\\
    &&\frac{\delta^2 W[C]}{\delta \sigma_{\mu\nu}(\theta_1)\,\delta \sigma_{\alpha\beta}(\theta_2)};\\
    && U_{a b} =\hat P \exp{ \int_a^b d \theta\, \dot C_\mu(\theta)\, A_\mu\!\big( C(\theta)\big)}\,.
\end{smalleq}
Choosing a ``hairpin'' loop $C_{\mathrm{HP}}$ with tips at $\mathbf r_1,\mathbf r_2$
and double strands between them (Fig.~\ref{fig::VPsi}) gives the standard gauge-invariant
two-point function:
\begin{smalleq}\label{HPformula}
     \VEV{\tr\!\Big[F_{\mu\nu}(r_1)\, U_{\theta_1,\theta_2}\,
          F_{\alpha\beta}(r_2)\, U_{\theta_2,\theta_1+2\pi}\Big]}
     = \left.\frac{\delta^2 W[C]}{\delta \sigma_{\mu\nu}(\theta_1)\,\delta \sigma_{\alpha\beta}(\theta_2)}\right|_{C=C_{\mathrm{HP}}}.
\end{smalleq}
\pctWPDF{0.5}{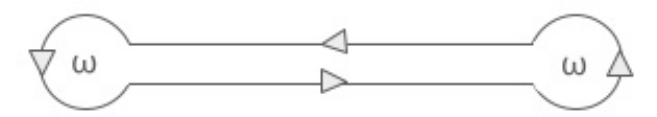}{The hairpin loop $C_{\mathrm{HP}}$ with small circles at $ r_1, r_2$ denoting
area derivatives, and two connecting strands between them. The symbol $\omega$ denotes the field strength.}

In the ``string'' language this correlator corresponds to a cylindrical worldsheet
bounded by two infinitesimal loops at $r_1$ and $r_2$, with two ``wires'' (Wilson
lines) drawn along the cylinder that connect the endpoints at $r_1$ and $r_2$.

Its large-distance asymptotics encodes the \emph{glueball spectrum}; the spin content
is selected by the projection tensor acting on the index pairs $(\mu\nu)$ and
$(\alpha\beta)$.
In particular, the mass 
m in a given channel is read from the large-distance fall-off $\exp{- m |r_1 - r_2|}$, which sets the physical length scale.

Higher-point, local gauge-invariant correlators are generated analogously by taking
higher area derivatives. Geometrically, the corresponding loop shrinks to a ``bike
wheel'' with multiple spokes (Wilson lines) joining the insertions (channels) on the
rim:
\pctWPDF{0.5}{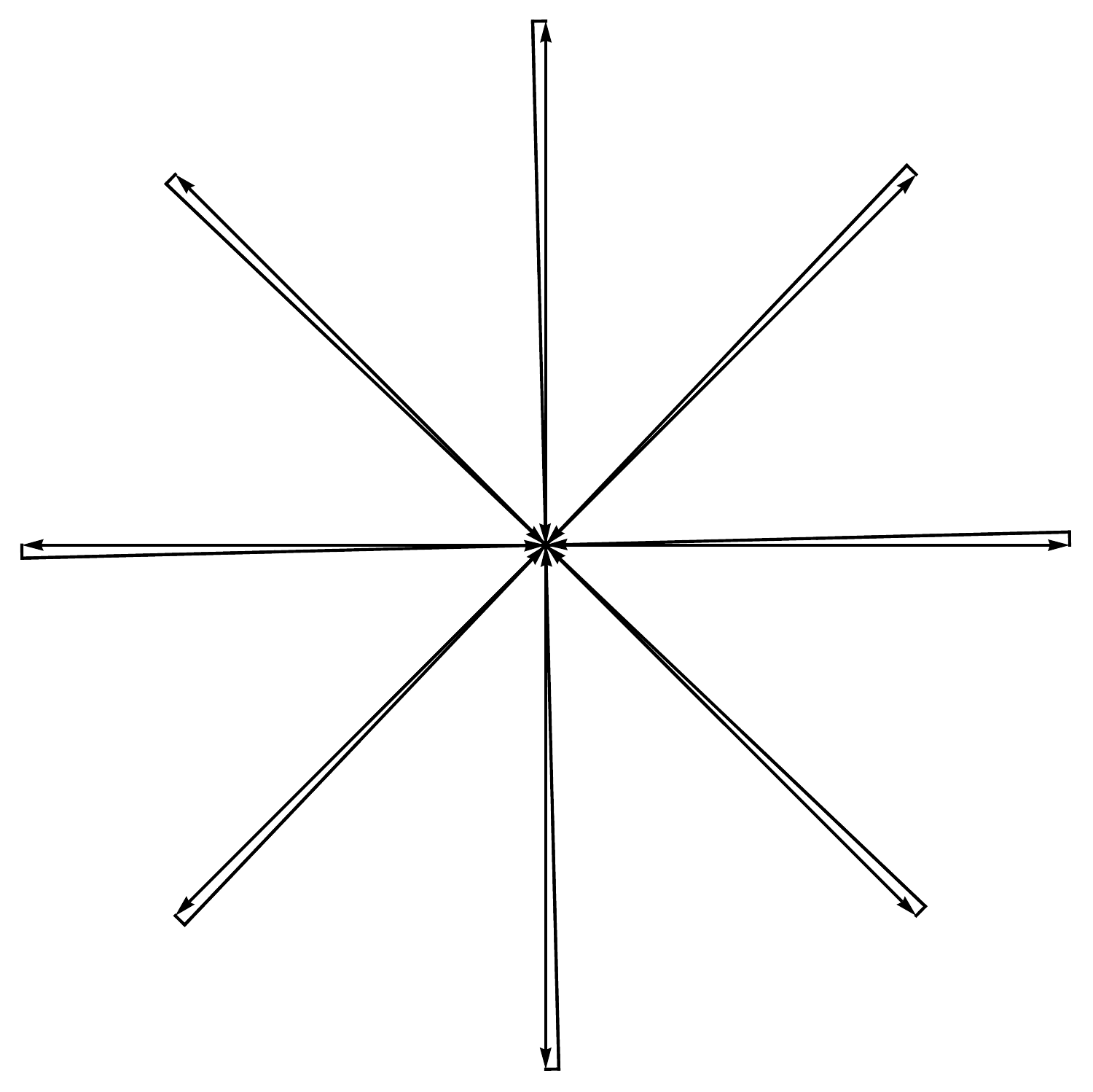}{A ``bike wheel'' loop generating multi-point, gauge-invariant
field-strength correlators in different glueball channels.}
We now reformulate the gradient-flow dynamics in loop space, where it becomes a
\emph{linear diffusion} amenable to exact solution in momentum-loop variables.
\hypertarget{sub-gradient-flow-as-loop-space-diffusion}{}%
\subsection{Gradient flow as loop--space diffusion}
\pdfbookmark[2]{Gradient flow as loop--space diffusion}{LoopCalculus}

The Wilson functional $W[C]$ \emph{probes} the gauge--field distribution and is linear in that measure. In the Abelian case,
\[
W[C] = \VEV{\exp{\I\,\Gamma_C}}, \qquad 
\Gamma_C = \oint_C A_\mu\,dx_\mu.
\]
In the non--Abelian case,
\[
W[C] = \VEV{\tr\,\hat P\,\exp{\oint \dot C\cdot A}},
\]
so the holonomy acts as a (noncommutative) characteristic functional of the circulation along $C$.

In loop space, the nonlinear PDE for $A(x,\tau)$ is replaced by a
\emph{linear diffusion--type} evolution:
\[
\partial_\tau W[C,\tau] = \mathcal{L}_C W[C,\tau],
\]
where $\mathcal{L}_C$ is the universal operator defined in Eq.~\eqref{LoopEq}, written as a line integral over the loop of a third--order functional derivative,
\[
\mathcal{L}_C \;:=\; \int d\theta\, \dot C(\theta)\cdot T \cdot 
\left(\frac{\delta}{\delta \dot C(\theta)}\right)^{\otimes 3}.
\]
The evolution is real and dissipative, with no imaginary unit.
The tensor $T$ is chosen so that, when applied to the ordered exponential of the gauge field, $\mathcal L_C$ reproduces the equation of motion $[D_\mu,F_{\mu\nu}]$. 
Moreover, $\mathcal{L}_C$ is linear in the strict sense of satisfying the Leibniz rules~\eqref{LeibnizRules1},~\eqref{LeibnizRules2}.

\emph{Universality.} The operator $\mathcal{L}_C$ is independent of 
the gauge--field measure; all dependence on $\mu[A(0)]$ lies in the 
initial condition $W[C,0]$. For $\tau>0$,
\[
W[C,\tau] = \exp{\tau\,\mathcal{L}_C}\,W[C,0].
\]
This same operator also governs hydrodynamic turbulence, with the 
velocity field in place of the gauge field. The crucial difference, as emphasized earlier, is that the hydrodynamic case is Abelian, whereas in Yang--Mills it is non--Abelian. This linear structure, together with the hidden symmetry of $\mathcal{L}_C$, makes asymptotic analysis near fixed--point manifolds or attractors \emph{analytically solvable}.

\hypertarget{sub-two-classes-of-turbulent-regimes}{}
\subsection{Two classes of turbulent regimes}
The key advantage of the diffusion picture is that turbulent regimes become \emph{classifiable and solvable} as linear problems with high internal symmetry.  
In the original (field) variables, turbulence is a nonlinear PDE problem with bifurcations and nontrivial probability measures on fields, often phenomenologically modeled by multifractals, but with spectra more subtle than simple power laws.  

In loop space, however, the turbulent problem becomes
a diffusion governed by $\mathcal{L}_C$ and is analytically solvable
\cite{migdal2023exact}: one studies the kernel (null space) of the loop--space
evolution operator,
\[
    (\partial_\tau-\mathcal{L}_C)\,W[C] \;=\; 0.
\]
This kernel has a rich number--theoretic structure (the Euler ensemble)
and carries a natural uniform measure --- precisely the manifestation
of spontaneous stochasticity in the original field description.
The Euler ensemble is universal, independent of initial data.

This turbulent solution corresponds to \emph{decaying turbulence} in loop--space diffusion:
a universal fixed trajectory rather than a fixed point.

A second class of universal turbulent solutions is the \emph{stationary turbulence}
associated with fixed points of the gradient flow.  
In terms of loop--space diffusion, this regime is described by the null space
of the diffusion operator itself,
\[
    \mathcal{L}_C\,W[C] \;=\; 0.
\]
Such a regime does not exist in pure hydrodynamics, where monotonic energy
dissipation drives the system to the trivial fixed point
$W[C,\infty]=1$, corresponding only to decaying turbulence.

In the present paper we find explicit solutions for both classes of turbulent
regimes in the \YM{} gradient flow.

\hypertarget{sub-the-evolution-of-the-wilson-loop-in-gradient-flow}{}
\subsection{The evolution of the Wilson loop in gradient flow}
Let us now derive the explicit evolution equation for the Wilson loop. 
This equation could have been written down forty years ago, after the loop calculus was first developed \cite{MMEq79, MM1981NPB, Mig83}. 

For readers unfamiliar with the original loop--calculus papers, we briefly review the essentials in the next section (see also \cite{M23PR,migdal2025duality}). In doing so, we advance the calculus by expressing variations in terms of the dot--velocity derivatives $\ff{\dot C(\theta)}$, which are finite and parametrization--invariant. This removes spurious singularities from the loop equation and enables clean checks of candidate solutions, as well as the construction of new ones.

We generalize the Wilson loop by replacing the averaging $\VEV{}_A$ over the \YM{} distribution by a more general averaging $\VEV{}_{A(\tau)}$ over the gradient flow starting from some (random or classical) initial field $A_\mu(0)$. In this case, the flow represents the evolution of the Wilson loop from its initial value, corresponding to the Yang--Mills quantum theory, or from any classical configuration. 

The derivation of the loop equation for a gradient flow starts from the identity following from the definition~\eqref{WCdef}:
\begin{smalleq}
   && \pd{\tau}  W[C,\tau] =  \int_0^{2\pi} d \theta\, \dot C_\nu(\theta)\, \nonumber\\
   && \VEV{\tr [D_\mu ,F_{\mu \nu}(C(\theta))]\,
        \hat P \exp{ \int_{\theta+}^{2 \pi + \theta-} d \theta'\,
        \dot C_\nu(\theta')\, A_\nu( C(\theta'))}}_{A(\tau)}.
\end{smalleq}
We set $\alpha=1$ by rescaling the stochastic time $\tau$.

\medskip
\noindent\textbf{Summary.}  
In this section we have reformulated the Wilson loop dynamics of the \YM{} gradient flow as a diffusion problem in loop space. This formulation reveals the universality of the operator $\mathcal{L}_C$, the linear and dissipative structure of the evolution, and the existence of two distinct turbulent regimes --- decaying and stationary. These features make the problem analytically solvable, in close parallel with the theory of decaying turbulence in hydrodynamics~\cite{migdal2023exact,migdal2024quantum}, and they provide the framework for constructing the exact solutions presented below.

\hypertarget{sec-the-loop-equations-without-singularities}{}
\section{The loop equations without singularities}
Before turning to the detailed derivations, we now translate the concepts 
outlined above into explicit mathematical form.  
We begin by formulating the classical planar QCD loop equation in the 
continuum, introducing the necessary loop-space calculus, and showing 
how the Hodge-dual minimal surface arises naturally as its exact 
fixed point solution.  
\hypertarget{sub-must-the-loop-equation-involve-singular-loops}{}
\subsection{Must the loop equation involve singular loops?}

Before we go into the details of this new calculus, we need to clarify a concern
that has persisted in discussions of the loop equation for decades.

In the original papers \cite{MMEq79, MM1981NPB, Mig83}, the area derivative and
loop gradient operations in the continuum loop equations were defined
geometrically by adding an infinitesimal closed loop $\delta C$ to the original
loop $C$ and then shifting this small loop in some direction. as shown on Figure \ref{fig::Loop_Calculus}.

\pctWPDF{0.5}{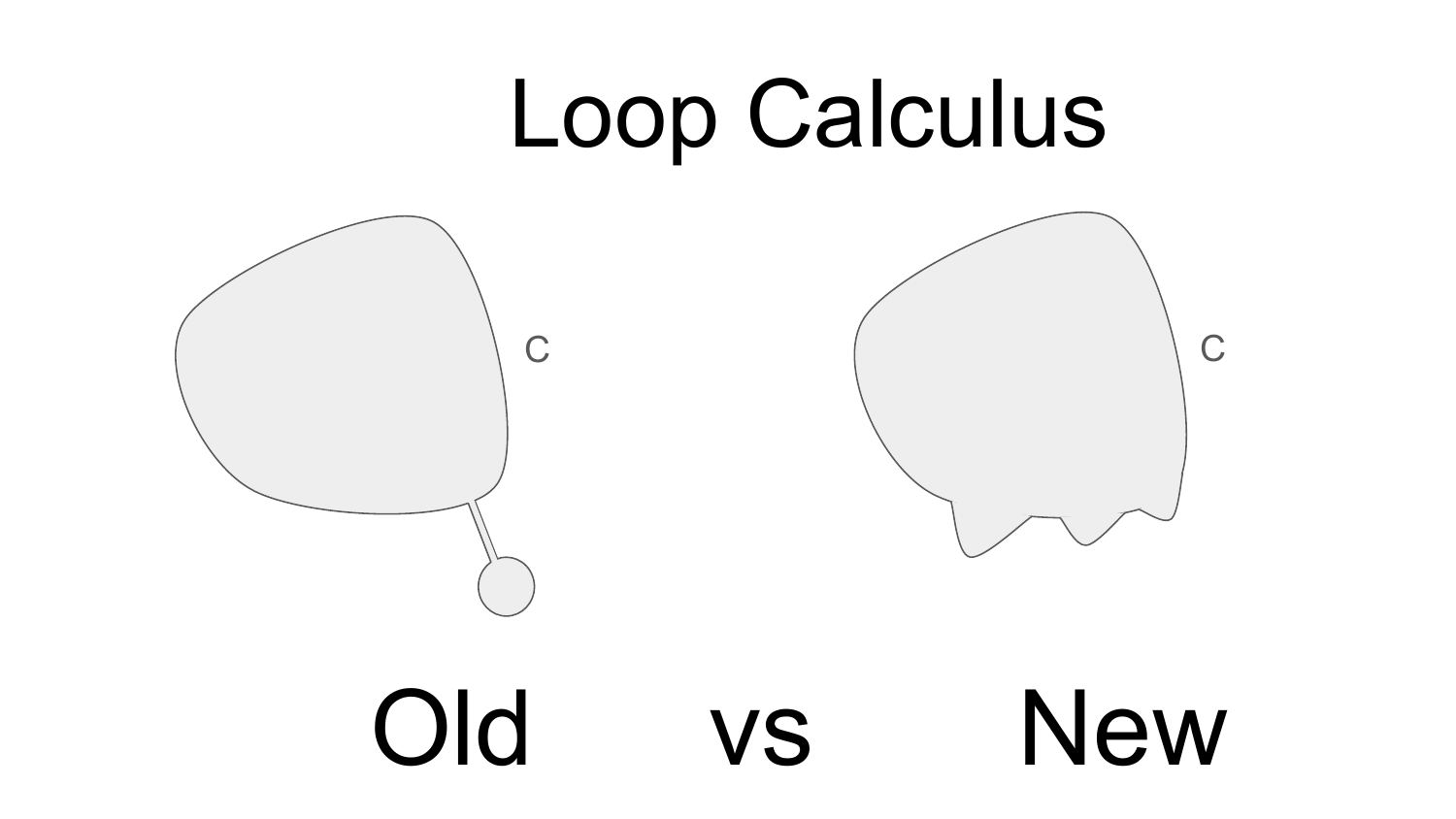}{The old loop equation involved adding an infinitesimal closed loop $\delta C$ near $C$ and connecting it with two little wires, creating a singular loop $\tilde{C} = C + \delta C + \mathrm{wires}$ that requires regularization and renormalization of the Wilson loop functional $W[\tilde{C}]$. In the new loop equation, only three smooth variations of the velocity $\dot{C}(\theta)$ are used, with their parameter positions $\theta\!-\!0$, $\theta$, and $\theta\!+\!0$ taken to coincide while preserving ordering. In this limit, a suitable linear combination of tensor components of the third functional derivative $\frac{\delta}{\delta \dot{C}_\alpha(\theta-0)}\frac{\delta}{\delta \dot{C}_\beta(\theta)}\frac{\delta}{\delta \dot{C}_\gamma(\theta+0)}$ produces the classical operator $D_\mu F_{\mu\nu}(C(\theta))$ when applied to the Wilson loop functional. The discontinuities of the third functional derivative as a function of $\theta\!-\!0$, $\theta$, and $\theta\!+\!0$ on the parameter circle are unrelated to singularities of quantum fields at coincident points in $\mathbb{R}^4$; they are kinematical effects of the ordering of covariant derivatives in the Wilson loop, and they cancel in the loop equation for Wilson loops built from classical fields obeying the equations of motion.
}
Similarly, in the lattice version of the loop equation, the small loop was the
size of the lattice spacing — one of the squares on an elementary cube.

When interpreted literally, these variations leave the space of smooth loops.
The addition of $\delta C$ necessarily creates a cusp or other non-smooth
feature at the point of attachment to $C$. In perturbative QCD, the evaluation
of $W[C+\delta C]$ for such a singular loop produces short-distance
divergences: planar graphs contained entirely within $\delta C$ give
contributions that scale as $\log|\delta C|$ rather than being proportional to
the area element $\delta\sigma_{\mu\nu}$. As a result, the area derivative in
its original definition does not exist without regularizing these short-distance
singularities. This is the singular-loop problem in its original form.

The resolution in the present work is straightforward: in the new loop
calculus, we avoid adding small loops or shifting them aside altogether.
Instead, the variational operators are defined entirely within the space of
smooth loops, using only smooth deformations of $C$. With this modification,
the classical part of the loop equation for smooth, non-intersecting loops is
completely well-defined and free from such singularities.

There are discontinuities in functional derivatives of the Wilson loop already
at the classical level, arising from the noncommutativity of covariant
derivatives of the gauge field. 
These discontinuities display as differences between the left and right limits of two 
or more such derivatives at coinciding points $\theta\pm 0$ at the parametric circle. 
They are present even for spatially constant gauge fields $A_\mu \in \mathfrak{su}(N)$,  
independent of potential short-distance singularities
in products of quantum gauge fields at coincident points. The latter manifest
for self-intersecting loops as the contact terms in the QCD loop equation\cite{MMEq79, MM1981NPB}, 
which do not exist in the \YM{} gradient flows, -- which makes the problem of solving the loop equation for such loops a well-defined classical problem.

One can apply the same loop operators to a given test functional of the loop
and check whether it obeys the loop equation; in particular, we can  verify the test solutions for the fixed points of the gradient flow.

There is, however, an inherent
ambiguity: any superposition of Wilson loops for gauge fields satisfying the
classical \YM{} equations will satisfy the fixed point of the loop equation. We have to select the solutions with proper general features, such as parametric, translational and rotational invariance and also proper behavior at infinity (later we discuss these properties in some detail).

However, if a test functional \textbf{does not} obey the fixed point loop equation, then it cannot be an asymptotic solution for this gradient flow. Therefore, while finding a correct solution is hard, rejecting an
incorrect candidate is straightforward.

As we shall later show, the exponential of minus the area of a conventional
minimal surface fails to obey the fixed point loop equation. Thus, the widely accepted Nambu–Goto string is
ruled out as an asymptotic solution for the Wilson loop at large loops.
\hypertarget{sub-the-loopcalculus}{}
\subsection{The \loopcalculus}\hypertarget{LoopCalculus}{}
\label{LoopCalculus}

We use the loop calculus developed in \cite{Mig83, Mig98Hidden} and reviewed
(with minor improvements) in \cite{migdal2025duality}. The \loopcalculus{} is
geometrically equivalent to the loop derivatives of
\cite{MMEq79, MM1981NPB, Mig83}, but it operates entirely within the manifold
of smooth loops.

By the \loopcalculus{} we mean a variational framework that replaces the
addition/translation of side loops by coincident limits of smooth velocity
variations $\delta\dot C(\theta\!-\!0)$, $\delta\dot C(\theta)$,
$\delta\dot C(\theta\!+\!0)$. In this way the loop itself remains smooth; the
classical part of the loop operator acting on $W[C]$ inserts the triple commutator
$\left[D_\mu,\left[D_\mu, D_\nu\right]\right]$ inside the trace of the ordered exponential, 
and no cusp/backtracking singularities are introduced.

\paragraph{Clarifying note (contact terms in QCD, not in gradient flow)}
For smooth, non-intersecting loops the quantum contact terms in QCD are absent, so the
loop equation becomes a clean classical variational problem, same as in gradient flow. 
For loops with self-intersections the contact terms are treated separately; their factorized
structure provides boundary conditions that remove the residual ambiguity
present for smooth loops.

Practically, the \loopcalculus{} supplies robust analytical tools to compute
loop operators on candidate functionals and to verify the loop equation without
regularization or renormalization ambiguities.

\hypertarget{sub-the-wilson-loop-and-non-abelian-stokes-theorem}{}
\subsection{The Wilson loop  and non-abelian Stokes theorem}
\pdfbookmark[2]{The Wilson loop and non-abelian Stokes theorem}{Crep}

The definition of the area derivative follows from the non-abelian version of the Stokes theorem for the variation of the Wilson loop
\begin{smalleq}
    \delta W[C] = \oint d \theta  \dot C_\mu(\theta) \delta C_{\nu}(\theta) \fbyf{W[C]}{\sigma_{\mu\nu}(\theta)}
\end{smalleq}
\hypertarget{sub-an-operator-identity}{}
\subsection{An operator identity}
We represent the loop $C$ as a point $x = C(0)$ plus the integral of the loop velocity $\dot C_\mu(\theta')$. Periodicity of $C(\theta)$ constrains the integral of velocity to be zero. We treat the Wilson loop as a function of $x$ and a functional of the velocity $\dot C_\mu(\theta)$.
\begin{smalleq}
\hypertarget{Crep}{}
\label{Crep}
    &&C_\mu(\theta) = x_\mu + \int_0^\theta d \theta' \dot C_\mu(\theta');\\
    &&  \int_0^{2\pi} d \theta' \dot C_\mu(\theta') =0;
 \end{smalleq}
 The way to derive the loop operators is to replace the Wilson loop by its operator representation
\begin{smalleq}\hypertarget{operatorIdentity}{}
\label{operatorIdentity}
     && W[C]\otimes I = \frac{1}{N} \VEV{\tr P \exp{\I \int_0^{2\pi} d \theta \dot C_\mu(\theta) D_\mu(x)}};\\
     && D_\mu(x) = \pd{x_\mu} + A_\mu(x)
\end{smalleq}
Here $I$ is the unit operator in the Hilbert space. As we shall shortly see, the right side is also proportional to the unit operator.
In this representation, every factor in the operator product
\begin{smalleq}
    && W[C] \otimes I = \frac{1}{N} \VEV{\tr\prod_{s=0}^{2\pi}\exp{\I d s \dot C_\mu(s)D_\mu(x)}}
\end{smalleq}
shifts arguments of the vector potential to the right $A_\mu(x) $  by $d C(\theta) = d \theta \dot{C}(\theta)$. As a result of disentangling operators
\begin{smalleq}
    &&\exp{\I d C_\mu(\theta) D_\mu(x)} =\nonumber\\
   &&\exp{ \I d C_\mu(\theta)\pd{x_\mu} } \exp{ \I d  C_\mu(\theta)A_\mu(x)} = \nonumber\\
    &&\exp{\I d  C_\mu(\theta)A_\mu(x + d C(\theta))}\exp{ \I d C_\mu(\theta)\pd{x_\mu} } 
\end{smalleq} 
we shift all arguments of the vector potentials along the loop and recover the conventional ordered exponent. 
The final translational operator equals unity
\begin{smalleq}
    \exp{ \I \left(\int_0^{ 2 \pi}d C_\mu(\theta)\right)\pd{x_\mu}} = \exp{ \I 0\pd{x_\mu}} =I
\end{smalleq}
due to the periodicity of the loop
\begin{smalleq}
    \int_0^{2 \pi}d C_\mu(\theta) = C_\mu(2\pi) -C_\mu(0) =0;
\end{smalleq}
Comparing the factors in front of the unit operator in the Hilbert space, we verify this representation of the Wilson loop.
In the following we shall skip factor $I$ by implying that Wilson loop is a c-number, despite of being represented as an operator in the Hilbert space.

In this operator form, the functional derivatives $\ff{\dot C_\mu(t\pm)}$ bring down the covariant derivative operator before or after the ordered product, which makes no difference due to the cyclic symmetry of the trace:
\begin{smalleq}\hypertarget{dotDerDmu}{}
\label{dotDerDmu}
   && \ff{\dot C_\mu(t\pm)}W(C(.),\tau) =\nonumber\\
   &&\VEV{\frac{1}{N}\tr D_\mu(C(t\pm))P \exp{\I \int_t^{t+2\pi} d s \dot C_\mu(\theta) D_\mu(C(t))}}
\end{smalleq}
In this formula, we used cyclic symmetry of the trace and invariance of the Wilson loop with respect to the choice of the origin on a circle $(0, 2 \pi) \Ra (t, t + 2 \pi)$.  
We call such functional derivatives \textbf{the dot derivatives}. 
\hypertarget{sub-ambiguity-of-the-dot-derivative-and-uniqueness-of-loop-operators}{}
\subsection{Ambiguity of the dot derivative and uniqueness of loop operators}
\pdfbookmark[2]{Ambiguity of the dot derivative and uniqueness of loop operators}{areader}

We need to clarify a subtle point.There is a formal $t$ dependence of the covariant derivative operator $D_\mu(x) = \pd{x} + A_\mu(x)$, taken at $x= C(t)$ which raises the question: should we treat the argument $ x = C(t)$ as a functional of the loop velocity $\dot C()$ ? Then we would have some more terms in functional derivative involving functional derivatives.
\begin{smalleq}
    \fbyf{C_\mu(t)}{\dot C_\nu(t\pm) } 
\end{smalleq}
If we define the loop as $\eqref{Crep}$ we get a constant term
\begin{smalleq}
    &&\fbyf{C_\mu(t)}{\dot C_\nu(t+) } = 0;\\
     &&\fbyf{C_\mu(t)}{\dot C_\nu(t-) } = \delta_{\mu\nu};
\end{smalleq}
These terms cancel in \eqref{dotDerDmu}, but there is a way to avoid these terms by first choosing $t=0$ when these terms are absent, and then taking advantage of cyclic symmetry of the trace and choosing  $t$ as the origin on the unit circle.

As a mathematical object, the dot derivative $\ff{\dot C(\theta)}$ is multivalued: it is defined modulo addition of a vector independent of $\theta$.
This can be seen from the formal relation of the dot derivative to the ordinary functional derivative for periodic function.
\begin{smalleq}\label{dotCffC}
    \ff{\dot C_\nu(\theta) } = -\int d \theta\ff{C_\nu(\theta) }
\end{smalleq}
This is an indefinite integral, defined modulo a constant addition. This indefinite constant, as we shall see, cancels in all the operators of the loop calculus, making them single-valued, as required.

\hypertarget{sub-new-representation-of-the-area-derivative}{}
\subsection{New representation of the area derivative}
Setting $t\pm \to t$ in \eqref{dotDerDmu} yields the same limit in this dot derivative of the Wilson loop.
However, these limits yield different results for the tensor of second dot derivatives, as two covariant derivatives will now be inserted in a chronological order.
The field strength (which is the commutator of two covariant derivatives) is generated by the antisymmetric part of this tensor:
\begin{smalleq}\hypertarget{areader}{}
\label{areader}
    &&\fbyf{W(C(.),\tau)}{\sigma_{\mu\nu}(t)} = \ff{\dot C_{\lb{\mu}}(t-)}\ff{\dot C_{\rb{\nu}}(t+)}W(C(.),\tau)  = \nonumber\\
    &&\VEV{\frac{1}{N}\tr F_{\mu\nu}(C(t))P \exp{\I \int_t^{t+2\pi} d s \dot C_\mu(\theta) D_\mu(C(t))}}
\end{smalleq}
In this equation and later throughout the paper, we use the notation $(\lb{\mu},\dots,\rb{\nu})$ for the antisymmetric part of the tensor
\begin{smalleq}
    X(\lb{\mu},\rb{\nu}) \equiv X(\mu,\nu) - X(\nu,\mu)
\end{smalleq}
\hypertarget{sub-antisymmetrization-not-the-commutator}{}
\subsection{Antisymmetrization, not the commutator}
\pdfbookmark[2]{Antisymmetrization, not the commutator}{annihil}

Note that this is \textbf{not the same} as the commutator
\begin{smalleq}
   && \ff{\dot C_{\lb{\mu}}(t-)}\ff{\dot C_{\rb{\nu}}(t+)}  =\nonumber\\
   &&\ff{\dot C_{\mu}(t-)}\ff{\dot C_{\nu}(t+)}-\ff{\dot C_{\nu}(t-)}\ff{\dot C_{\mu}(t+)};\\
   &&\left[\ff{\dot C_{\mu}(t-)},\ff{\dot C_{\nu}(t+)}\right] = \nonumber\\
   && \ff{\dot C_{\mu}(t-)}\ff{\dot C_{\nu}(t+)}-\ff{\dot C_{\nu}(t+)}\ff{\dot C_{\mu}(t-)}
\end{smalleq}
The commutator of any functional derivatives yields zero, as it is well known.  In other words, the second functional derivative is symmetric with respect to the exchange of both indices and arguments $ (\mu, t-) \leftrightarrow (\nu,t+)$.  Therefore, our interchange of just indices $ \mu \leftrightarrow \nu$ is equivalent to the exchange of the arguments $ t- \leftrightarrow t+$.

This exchange of arguments of functional derivatives switches the order of operators of the covariant derivatives $D_\mu(C(t)) D_\nu(C(t))$ inside the trace of the path-ordered product, which creates the commutator:
\begin{smalleq}
    &&\left(\ff{\dot C_{\mu}(t-)}\ff{\dot C_{\nu}(t+)}-\ff{\dot C_{\mu}(t+)}\ff{\dot C_{\nu}(t-)}\right)\nonumber\\
   &&\VEV{\frac{1}{N}\tr P \exp{\I \int_t^{t+2\pi} d \theta\dot C_\mu(\theta) D_\mu(C(t))}} =\nonumber\\
    &&\frac{1}{N}\VEV{\tr \left[D_{\mu}(C(t)),D_{\nu}(C(t))\right]\right.\nonumber\\
   &&\left.P \exp{\I \int_t^{t+2\pi} d \theta \dot C_\mu(\theta) D_\mu(C(t))}}
\end{smalleq}
This is how we represent the non-commuting operators inside the trace by using antisymmetric tensor part of the ordinary  commuting product of functional derivatives.

The covariant derivative of $F_{\mu\nu}$ is generated by the difference of left and right dot derivatives
\begin{smalleq}
   && \partial_\mu(t)= \ff{\dot C_\mu(t-)}-\ff{\dot C_\mu(t+)};\\
   &&\partial_\mu(t) \fbyf{W[C]}{\sigma_{\mu\nu}(t)} = \nonumber\\
   &&\VEV{\tr \left[D_\mu, F_{\mu\nu}(C(t))\right]P \exp{\I \int_t^{t+2\pi} d \theta \dot C_\mu(\theta) D_\mu(C(t))}}
\end{smalleq}
\hypertarget{sub-parametric-invariance}{}
\subsection{Parametric invariance}
Note that, unlike the ordinary functional derivative, the dot derivative is parametric invariant $ s\Ra f(s), f'(s) >0$, as it follows from the definition
\begin{smalleq}
    \delta F[C] = \int d s \delta \dot C_\mu(s) \fbyf{F}{ \dot C_\mu(s)}
\end{smalleq}
It also follows from this definition that this functional derivative is defined modulo an addition of a constant term  $\int d s \delta \dot C(s) \const{}=0$ by periodicity of $C(s)$.
As we see, the left and right dot derivatives $\ff{ \dot C_\gamma(\theta\pm 0)}$ must produce different limits for the area and point derivatives to be finite. 
These functional derivatives, unlike conventional $\ff{C_\mu(s)}$, do not produce delta functions, but rather finite results which may differ for the right and left limit at every point. This is clear from the above formulas for the area derivative and the covariant derivative of the field strength.

\hypertarget{sub-annihilation-of-the-wilson-loop}{}
\subsection{Annihilation of the Wilson loop}
Another important property of the dot derivative is that its discontinuity $\partial_\mu(t)$ annihilates the Wilson loop
\begin{smalleq}\hypertarget{annihil}{}
\label{annihil}
    \partial_\mu(t) W(C(.),\tau) =0;
\end{smalleq} 
There are many ways to check it, but the simplest one is to represent 
\begin{smalleq}
    \partial_\nu(t) = - \int_{t-}^{t+} d s \pd{s} \ff{\dot C_\nu(s)} = \int_{t-}^{t+} d s  \ff{C_\nu(s)} 
\end{smalleq}
Now, using the basic relation
\begin{smalleq}\label{fCffsigma}
    \fbyf{W[C,\tau]}{C_{\nu}(s)} =  \dot C_\mu(s)  \fbyf{W[C,\tau]}{\sigma_{\mu\nu}(s)}
\end{smalleq}
we conclude that with a finite area derivative, the integral 
\begin{smalleq}
    \partial_\nu(t) W(C(.),\tau)  = \int_{t-}^{t+} d s  \dot C_\mu(s)  \fbyf{W[C,\tau]}{\sigma_{\mu\nu}(s)}=0
\end{smalleq}
This is true for the Wilson loop both in the Abelian and the non-Abelian cases.

As we mentioned above, the dot derivative is defined modulo the addition of an arbitrary constant vector. This makes its discontinuity $\partial_\mu(t)$ uniquely defined. The antisymmetric product of the left and right dot derivatives in the definition of the area derivative of the Wilson loop is also unique, due to the annihilation property \eqref{annihil}. Shifting both dot derivatives by the same constant $\ff{\dot C_\mu(t\pm)} \Ra \ff{\dot C_\mu(t\pm)} + a_\mu$ we find that the extra term in the antisymmetric product \eqref{areader} is proportional to \eqref{annihil} and thus it vanishes.

 Likewise, one can prove that the area derivative annihilates any local function of $C(\theta)$.
\begin{smalleq}
    &&\ff{\dot C_{\lb{\mu}}(t-)}\ff{\dot C_{\rb{\nu}}(t+)}F(C(\theta_1)) = \nonumber\\
   &&\ff{\dot C_{\lb{\mu}}(t-)}\ff{\dot C_{\rb{\nu}}(t+)} F\left(\int^{\theta_1}d \theta_2  \dot C(\theta_2)\right) \nonumber\\
    &&\propto\ff{\dot C_{\lb{\mu}}(t-)} \pd{\rb{\nu}} F(C(\theta_1))\Theta((t+)-\theta_1)\nonumber\\
    &&\propto \pd{\lb{\mu}} \pd{\rb{\nu}} F(C(\theta_1))\Theta((t+) -\theta_1)\Theta(\theta_1 -(t-)) =0
\end{smalleq}
The antisymmetric product of two gradient operators yields zero by definition.
Thus, we conclude that for any local function of a vector $F(r)$
\begin{smalleq}\hypertarget{noareader}{}
\label{noareader}
    \fbyf{F( C(\theta_1))}{\sigma_{\mu\nu}(\theta)} = 0 \;\forall \theta, \theta_1
\end{smalleq}
\hypertarget{sub-the-leibniz-rules}{}
\subsection{The Leibniz rules}
\pdfbookmark[2]{The Leibniz rules}{LeibnizRules1}

As it was noted already in the first papers \cite{MMEq79, MM1981NPB}, the area derivative, though formally being built from the second functional derivative, acts as a first order functional derivative on the Stokes-type functionals. To be specific, it satisfies the Leibniz rules 
\begin{smalleq}\hypertarget{LeibnizRules1}{}
\label{LeibnizRules1}
   && \ff{\sigma_{\mu\nu}(\theta)} (A[C] B[C]) = \fbyf{A[C]}{\sigma_{\mu\nu}  (\theta)} B[C] + A[C] \fbyf{B[C]}{\sigma_{\mu\nu}  (\theta)};\\
   && \ff{\sigma_{\mu\nu}(\theta)} F(A[C]) = F'(A[C]) \fbyf{A[C]}{\sigma_{\mu\nu}  (\theta)}
\end{smalleq}
In our \loopcalculus, with the definition of area derivative as antisymmetrized product of two dot derivatives, these rules follow from the annihilation of the Stokes-type functional by derivative operator \eqref{annihil}.

The Leibniz rule is very important in solving the loop equation. In particular, the second rule plus annihilation rule \eqref{annihil}, applied to the exponential of the Hodge-dual minimal area ( see below) immediately leads to the conclusion that self-duality of this surface makes this exponential a solution of the fixed point loop equation.

The first Leibniz rule allows multiplying two functionals, each solving the fixed point loop equation. The product also satisfies this equation
\begin{smalleq}\hypertarget{LeibnizRules2}{}
\label{LeibnizRules2}
   &&\pd{\mu} \ff{\sigma_{\mu\nu}(\theta)} (A[C] B[C]) =\nonumber\\
   &&\pd{\mu}\fbyf{A[C]}{\sigma_{\mu\nu}  (\theta)} B[C] + A[C] \pd{\mu}\fbyf{B[C]}{\sigma_{\mu\nu}  (\theta)} =0;
\end{smalleq}
Here we also used the annihilation rule \eqref{annihil}.
\hypertarget{sub-the-bianchi-identity-and-the-loop-equation}{}
\subsection{The Bianchi identity and the Loop equation}
\pdfbookmark[2]{The Bianchi identity and the Loop equation}{Bianchi}
\hypertarget{Bianchi}{}
\label{Bianchi}
There are two consequences from the above representation of the loop derivatives in terms of the dot derivatives. First, there is a Bianchi identity
\begin{smalleq}
e_{\alpha\mu\nu\lambda}\partial_\alpha(\theta)\fbyf{W[C,\tau]}{\sigma_{\mu\nu}(\theta)} = 0
\end{smalleq}
which holds at the kinematical level, regardless of dynamics.
This identity holds on the operator level for the dot derivatives
\begin{smalleq}\hypertarget{ident}{}
\label{ident}
   && e_{\alpha\mu\nu\lambda} \left(\ff{\dot C_\alpha(\theta+)} - \ff{\dot C_\alpha(\theta-)}\right) \ff{\dot C_{\lb{\nu}}(\theta-)} \ff{\dot C_{\rb{\mu}}(\theta+)} =0;
\end{smalleq}
\hypertarget{sub-the-nonsingular-loop-equation}{}
\subsection{The nonsingular loop equation}
\pdfbookmark[2]{The nonsingular loop equation}{LoopEq}

Second, we find the evolution of the Wilson loop in the gradient flow
\begin{smalleq}\hypertarget{LoopEq}{}
\label{LoopEq}
    && \pd{\tau}  W[C,\tau] = \int_0^{2\pi} d \theta \dot C_\nu(\theta) \left(\ff{\dot C_\mu(\theta+)} - \ff{\dot C_\mu(\theta-)}\right)\nonumber\\
   &&\ff{\dot C_{\lb{\nu}}(\theta-)} \ff{\dot C_{\rb{\mu}}(\theta+)}W[C,\tau];
\end{smalleq}
This can also be written as diffusion equation in loop space:
\begin{smalleq}\hypertarget{LoopEq}{}
\label{LoopEq}
    && \pd{\tau}  W[C,\tau] = \mathcal L_C W[C,\tau];
\end{smalleq}
with diffusion operator
\begin{smalleq}\hypertarget{diffusionOperator}{}
\label{diffusionOperator}
   && \mathcal L_C = \oint d \theta \dot C_\nu(\theta) \hat L_\nu(\theta);\\
   &&\hat L_\nu(\theta) = T^{\alpha\beta\gamma}_\nu \frac{\delta^3}{\delta \dot C_\alpha(\theta-0)\delta \dot C_\beta(\theta)\delta \dot C_\gamma( \theta+0)};\\
   &&T^{\alpha\beta\gamma}_\nu =\delta_{\alpha\beta}\delta_{\gamma\nu}+ \delta_{\gamma\beta}\delta_{\alpha\nu}-2\delta_{\alpha\gamma}\delta_{\beta\nu};
\end{smalleq}
All three arguments of the third functional derivative tend to $\theta$ in specified order. The dot derivatives generate covariant derivatives inside the ordered product in the Wilson loop, and contraction with the tensor the tensor $T$ arranges these operators into a triple commutator 
\begin{smalleq}
    T^{\alpha\beta\gamma}_\nu D_\alpha D_\beta D_\gamma =\left[D_\mu,\left[D_\mu,D_\nu\right]\right];
\end{smalleq}
\hypertarget{sub-symmetric-form-of-loop-space-diffusion}{}
\subsection{Symmetric form and null space of loop space diffusion operator}
\pdfbookmark[2]{Symmetric form and null space of loop space diffusion operator}{symForm}

This subsection elaborates on the \emph{trilinear (local)} form of the diffusion operator introduced just above by exhibiting an equivalent square–of–gradient representation and identifying the source of its nontrivial null space.

In its original formulation, the \YM{} gradient flow is defined by the time derivative of the gauge field being equal to minus the variation of a positive definite energy functional. In this form, dissipation is manifest: the energy monotonically decreases, so the system converges to a classical fixed point.

Starting \emph{from the trilinear local representation} of $\mathcal L_C$, we use the relation between the dot derivative and the ordinary functional derivative to write the jump across $\theta$ as an ordered loop integral (the “opposite direction” traversal):
\begin{smalleq}
  && \ff{\dot C_\mu(\theta{+})} - \ff{\dot C_\mu(\theta{-})}
     \;=\; \int_{\theta}^{\theta + 2\pi} d\theta'\, \ff{C_\mu(\theta')} \,,
\end{smalleq}
and, using \eqref{fCffsigma},
\begin{smalleq}
  && \ff{\dot C_\mu(\theta{+})} - \ff{\dot C_\mu(\theta{-})}
     \;=\; \int_{\theta}^{\theta + 2\pi} d\theta'\, \dot C_\alpha(\theta')\, \ff{\sigma_{\alpha\mu}(\theta')} \,.
\end{smalleq}
Substituting this identity into the trilinear form yields the symmetric “square–of–gradient” representation
\begin{smalleq}
  && \mathcal L_C \;=\; \oint d\theta\, \mathcal R_\mu(\theta)\,
        \int_{\theta}^{\theta + 2\pi} d\theta'\, \mathcal R_\mu(\theta') \,;\\
  && \mathcal R_\mu(\theta) \;=\;  \ff{C_\mu(\theta)}  \;=\; \dot C_\alpha(\theta)\, \ff{\sigma_{\alpha\mu}(\theta)} \,.
\end{smalleq}
Here $\mathcal R_\mu(\theta)$ plays the role of a gradient in loop space; since the area derivative is second order \eqref{areader}, $\mathcal R_\mu(\theta)$ is second order and $\mathcal L_C$ is fourth order. This form is manifestly symmetric and positive semidefinite.

\emph{Ordered vs.\ full integral and the null space.}
The ordered integral $\int_{\theta}^{\theta + 2\pi} d\theta'\, \mathcal R_\mu(\theta')$ explicitly \emph{excludes} the diagonal point $\theta'=\theta$. For reparametrization/translation–invariant functionals (e.g.\ Wilson loops) the complete integral vanishes,
\begin{smalleq}
  && \oint d\theta\, \mathcal R_\mu(\theta)\, W[C]
   \;=\; \oint d\theta\, \dot C_\alpha(\theta)\, \ff{\sigma_{\alpha\mu}(\theta)} W[C]\nonumber\\
   &&=\; \oint d\theta\, \ff{C_\mu(\theta)} W[C]
   \;=\; 0 \,,
\end{smalleq}
so the nonlocal quadratic form is blind to the constant-in-$\theta$ mode. The \emph{missing diagonal} is distributional (contact term $\sim \delta(\theta'-\theta)$); reinstating it reproduces precisely the local third–order (trilinear) operator derived above. Thus the nontrivial null space of $\mathcal L_C$ does \emph{not} come from a pointwise kernel $\mathcal R_\mu(\theta)W[C]=0$, but from the projection induced by the ordered double integral that omits the diagonal. This is the loop–space analogue of a degenerate diffusion \cite{Montgomery2002} (positive semidefinite with a structured kernel), in sharp contrast to a strictly elliptic Laplacian in Euclidean space.

In what follows we will continue to use the trilinear local form for verification of exact solutions; the symmetric view is recorded here only to clarify the origin of the null space.
\hypertarget{sub-minimal-area-and-its-area-derivative}{}
\subsection{Minimal area and its area derivative}
\pdfbookmark[2]{Minimal area and its area derivative}{innerNormal}

We review, for readers' convenience, in appendix.1  the classical theory of minimal surfaces in a form we need in this paper.
This review follows the lines of the chapter "Minimal Surfaces" in \cite{M23PR}.

The area of the parametric surface $S: x = X(\xi)\in \mathbb R^4, \xi = (\xi_1,\xi_2)$  is defined by an area element
\begin{smalleq}
    d \Sigma_{\mu\nu} = 1/2 \tr d X_\mu \wedge d X_\nu
\end{smalleq}
In components, it reads
\begin{smalleq}
  &&  d \Sigma_{\mu\nu}(\xi) = d ^2\xi  \sigma_{\mu\nu}(\xi);\\
    && \sigma_{\mu\nu}(\xi) = e_{l m}\pd{l} X_\mu(\xi) \pd{m} X_\nu(\xi)
\end{smalleq}
We presume the topology of a disk with polar coordinates $\xi = (r,\theta), r  \in (0,1), \theta\in (0,2 \pi)$. The boundary loop $C$ corresponds to the Dirichlet boundary condition.
\begin{smalleq}
    X_\mu(1,\theta) = C_\mu(\theta)
\end{smalleq}
The surface area is defined as the integral of the absolute value of the area element
\begin{smalleq}
  &&  |S| =\int \abs{d \Sigma_{\mu\nu}} = \int d^2 \xi \sqrt{\oh \sigma_{\mu\nu}^2};
\end{smalleq}
The corresponding area variation ( under variation of $X_\mu(\xi)$)
\begin{smalleq}
    && \delta |S| = \int d^2 \xi \delta \sigma_{\mu\nu} t_{\mu\nu};\\
    && \delta \sigma_{\mu\nu} = 2 e_{l m}\pd{l}\delta X_\mu(\xi) \pd{m}  X_\nu(\xi);\\
    && t_{\mu\nu} = \frac{ \sigma_{\mu\nu}}{\sqrt{\oh\sigma_{\alpha\beta}^2}}
\end{smalleq}
This tensor $t_{\mu\nu}$ is a normalized normal tensor.
The extremality condition leads to the Euler-Lagrange equation
\begin{smalleq}
    \fbyf{|S|}{X_\mu(\xi)} =
    -2 e_{l m} \pd{l} t_{\mu\nu} \pd{m} X_\nu=0
\end{smalleq}
These equations are studied in minimal surface theory, which originated with Plateau and Weierstrass (we summarize it in appendix.1).
The  variation of the minimal area by a boundary curve follows from the extremality conditions and Stokes theorem
\begin{smalleq}
   && \delta|S|_C = \int  d^2 \xi\delta \sigma_{\mu\nu}t_{\mu\nu} =\nonumber\\
   &&\int d^2 \xi \left(\cancel{-2 e_{l m} \pd{l} t_{\mu\nu} \pd{m} X_\nu}\right) + 
   2 \oint d \theta t_{\mu\nu}(1,\theta) \delta C_\mu(\theta) \dot C_\nu(\theta)=\nonumber\\
   && 2 \oint d \theta t_{\mu\nu}(1,\theta) \delta C_\mu(\theta) \dot C_\nu(\theta)
\end{smalleq}
from which we find
\begin{smalleq}
   \fbyf{|S|}{\sigma_{\mu\nu}(\theta)} = 2 t_{\mu\nu}(1,\theta);
\end{smalleq}
\hypertarget{sub-violation-of-the-loop-equation-by-the-minimal-area}{}
\subsection{Violation of the loop equation by the minimal area}

To test whether a given surface satisfies the fixed point loop equation, we apply a loop operator to the area functional and examine whether the result vanishes. This loop operator probes the local sensitivity of the surface area to variations in the tangent vector of the boundary loop.

We consider the Weierstrass-Douglas minimal surface \cite{Osserman2002} constructed as the harmonic extension of a boundary loop $C^\mu(t)$, where $t$ parametrizes the unit circle by conformal mapping of a unit disk onto the upper semiplane. The real part of the Cauchy-type singular integral gives the tangent surface vector $N^\mu(t)$ normal to the boundary tangent vector  $\dot C$. Up to irrelevant normalization
\begin{smalleq}\hypertarget{innerNormal}{}
\label{innerNormal}
N^\mu(t) = \text{Re} \int_{=\infty}^\infty \, df \frac{\dot C^\mu(f)}{\tau(f) - \tau(t) - i\varepsilon} .
\end{smalleq}
where $\tau(f)$ is parametrization of the boundary, to be determined in turn by minimization of the Douglas functional \eqref{Douglas} by an inverse function $f(\tau)$. The implicit nonlocal relation between $\tau$ and $f$ is given by \eqref{parametrization}.

We derive this and other formulas of Weierstrass-Douglas theory in appendix.1
The area derivative of this minimal area, as we derived in the previous section, is a tensor area element at the boundary $C$ of the minimal surface. This is a bivector formed from the tangent vector and the normal vector:
\begin{smalleq}
&&\frac{\delta A}{\delta \sigma_{\mu\nu}(t)}  = 2 \frac{\dot C_{\lb{\mu}}(t) N_{\rb{\nu}}(t)}{Z};\\
&& Z = \sqrt{\left(C_{\lb{\mu}}(t) N_{\rb{\nu}}(t)\right)^2}
\end{smalleq}

We now apply the derivative operator $\pd{\mu}$, which consists of taking the functional derivative of this normalized bivector with respect to $\dot C^\rho(t')$, and then extracting the discontinuity of the result across $t' = t$.

This functional derivative produces two terms:

\begin{enumerate}
\item The first term arises from varying $\dot C^\mu(t)$ directly:
\begin{smalleq}
\frac{\delta \dot C^\mu(t)}{\delta \dot C^\rho(t')} = \delta^\mu_\rho \, \delta(t - t').
\end{smalleq}
Since $\delta(t - t')$ is an even distribution, its discontinuity vanishes:
\begin{smalleq}
\text{Disc}[\delta(t - t')] = 0.
\end{smalleq}
To make this evenness explicit, recall the well-known representation of the delta function:
\begin{smalleq}
\pi \delta(a) = \lim_{\varepsilon \to 0} \frac{\varepsilon}{a^2 + \varepsilon^2}.
\end{smalleq}
This function is symmetric in $a$, so its left and right limits agree for any fixed $\varepsilon > 0$, and hence its discontinuity is identically zero.
Thus, this term does not contribute to the loop operator.

\item The second term arises from varying the normal vector:
\begin{smalleq}\hypertarget{singterm}{}
\label{singterm}
\dot C_{\lb{\mu}}\frac{\delta N_{\rb{\nu}}(t)}{\delta \dot C^\mu(t')} = -3 C_\nu \, \text{Re} \left( \frac{1}{\tau(t') - \tau(t) - i\varepsilon} \right).
\end{smalleq}
This kernel is an odd function of $\delta = \tau'(t)(t' - t)$ in the vicinity of $\delta=0$, and its discontinuity is:
\begin{smalleq}\label{deltaEpslimit}
\left[\frac{\delta}{\delta^2 + \varepsilon^2}\right]^{\delta}_{-\delta} = \frac{2\delta}{\delta^2 + \varepsilon^2}.
\end{smalleq}
As both $\delta \to 0$ and $\varepsilon \to 0$, this expression diverges and becomes indefinite, depending on the limiting path. Therefore, it does not vanish.
\end{enumerate}

We conclude that the result of applying the loop operator to the Douglas minimal surface is nonzero and singular. The fixed point loop equation, which demands a smooth and vanishing result, is violated. This failure is not due to quantum effects or contact terms but is a structural feature of the surface itself.
\noindent\textit{Clarification.} The classical gradient flow contains \emph{no} contact terms; such terms arise only in the Langevin (stochastic) quantization of QCD and are tuned to zero in the local limit considered here.

It is worth remembering at this point, that the fixed point loop equation comes from the triple product of covariant derivatives applied to the Yang--Mills field. There are no divergencies in these classical equations when applied to differentiable fields. We may expect the same for the smooth functionals like a classical minimal area.  The singular nonvanishing result means that the classical minimal surface does not satisfy the fixed point loop equations. 
\subsection{Previous attempts to satisfy the loop equation by minimal area}

The AdS/CFT conjecture \cite{Maldacena1998} inspired many analyses
\cite{Drukker_Gross_Ooguri1999,PolyakovRychkov} that sought nonperturbative
evidence by testing whether \emph{minimal surfaces} satisfy loop equations for
appropriate classes of loops. How does our result compare with these works?

First, one should separate two settings: (i) supersymmetric Yang--Mills (SYM) and
its conjectured dual to string theory in AdS, and (ii) \emph{pure} Yang--Mills and a
putative dual description by minimal areas in \emph{four-dimensional Euclidean}
space. In this paper we work strictly in the latter setting: we study the loop
equation in \emph{Euclidean 4-space} and test whether the \emph{Euclidean} minimal
surface with the same boundary satisfies that equation. We therefore do not take a
position (here) on the AdS/CFT minimal-area prescriptions for circular superloops
\cite{Drukker_Gross_Ooguri1999}; our analysis neither proves nor disproves those
claims, though we expect consistency within that supersymmetric context.

Regarding \cite{PolyakovRychkov}, there is partial overlap in spirit but also a
clear point of divergence. Setting aside their AdS minimal surfaces as not directly
applicable to our Euclidean setup, the apparent tension is that with certain
smoothness assumptions they conclude that minimal area satisfies a scalar (Laplace)
version of the loop equation, whereas we find---with a finite, nonsingular loop
operator and no auxiliary smoothness assumptions---that the \emph{Euclidean} minimal
area \emph{does not} satisfy the full \YM{} loop equation.

The main technical differences are as follows.
\begin{itemize}
  \item \textbf{Finite vs.\ singular loop operator.}
  We employ exact identities for Stokes-type functionals and gauge fields to
  formulate a \emph{finite} loop equation (removing spurious delta function terms). The
  ``loop-space Laplacian'' used in \cite{PolyakovRychkov} is singular and can mix
  with perturbative short-distance divergences in asymptotically free QCD (effects
  that are absent in our classical-flow setting).

  \item \textbf{Full \YM{} content vs.\ scalar projection.}
  Our equation uses the \emph{full} \YM{} structure: the four Euler--Lagrange
  components together with the four Bianchi components (nontrivial in loop space).
  In \cite{PolyakovRychkov} a \emph{scalar} (Laplace) projection is analyzed and a
  ``zigzag'' symmetry is imposed to restrict to Stokes-type functionals. In our
  framework, Stokes-ness follows from the Bianchi identity and the area derivative,
  and we show that the minimal area is indeed Stokes-type---yet it still fails the
  full \YM{} loop equation.

  \item \textbf{Classical gradient flow vs.\ quantum/UV issues.}
  We consider \emph{classical} \YM{} gradient flow (no Langevin forcing), so there
  are no perturbative UV divergences to disentangle and no need for OPE/conformal
  tools. For nonselfintersecting loops, the loop equation is free of such
  singularities even in the quantum theory.

\item \textbf{Self-intersections not required for the negative result.}
While any positive proof of a QCD area law would ultimately have to treat
self-intersecting loops, our \emph{negative} conclusion for the Euclidean minimal
area already follows from \emph{nonintersecting} contours.

  \item \textbf{Exact minimal surface vs.\ ``wavy'' approximation.}
  We use the \emph{exact} Weierstrass representation of the Euclidean minimal
  surface and substitute it into the \emph{finite} loop equation. In
  \cite{PolyakovRychkov} exact minimal surfaces are not used; instead one analyzes
  sufficiently smooth, ``wavy'' loops.

  \item \textbf{Path dependence at the boundary.}
  For the Euclidean minimal area our finite loop operator is \emph{nonzero} and, in
  fact, \emph{singular}: its value depends on the path along the minimal surface
  approaching the boundary loop. The positive claim in \cite{PolyakovRychkov} is
  consistent with our restricted limit $\delta\ll\epsilon$ in \eqref{deltaEpslimit},
  but this does not cover the generic situation.

  \item \textbf{A different (higher-dimensional) minimal surface does work.}
  There exists another minimal surface---in a sixteen-dimensional ambient space with
  four-dimensional boundary---that \emph{does} satisfy the same equation without
  singularities or ambiguities. We construct it later in the paper.
\end{itemize}

\hypertarget{sub-could-the-ordinary-minimal-surface-be-rescued-by-subleading-terms}{}

\subsection{Could the ordinary minimal surface be rescued by subleading terms?}
\pdfbookmark[2]{Could the ordinary minimal surface be rescued by subleading terms?}{LoopArea}

It has been suggested that the failure of the ordinary minimal surface in $\mathbb{R}^4$ to satisfy the fixed--point loop equation might be cured by adding subleading (large--area suppressed) corrections to the minimal--area functional.  
For QCD, several structural obstacles make this route implausible; in the Yang--Mills gradient--flow setting these issues do not arise, since there are no contact terms and hence no short--distance singularities.  
This subsection is included for readers interested in possible QCD applications or in the broader question of \emph{spontaneous quantization}.

\paragraph{Scaling mismatch of singularities.}  
For a smooth loop of size $L$ with minimal separation $a$ (serving as UV cutoff), the loop operator applied to the area--law ansatz $W[C] = \exp{-\sigma\, \mathrm{Area}}$ produces in~\eqref{singterm} a contribution
\begin{smalleq}\hypertarget{LoopArea}{}
\label{LoopArea}
   \frac{\sigma L}{a}.
\end{smalleq}
In perturbative QCD, by contrast, the loop equation contains a contact term from one--gluon exchange with the propagator replaced by its Laplacian. For a smooth loop this gives
\begin{smalleq}\hypertarget{contactTerms}{}
\label{contactTerms}
   \frac{L}{a^3 \log a},
\end{smalleq}
which is parametrically more singular. Any attempt to cancel such a term by subleading large--area corrections would require unnatural fine--tuning across different orders of singularity.

\paragraph{Two--cutoff problem and directional dependence.}  
For the Weierstrass minimal surface, the singular variation of the normal vector depends on two independent regulators:  
$\Delta\theta$ (parametric separation along the loop) and $\delta z = \eps$ (distance from inside the surface, i.e.\ the imaginary shift in the upper half--plane).  
The loop--equation contribution depends on their ratio $\Delta\theta/\eps$, an unphysical parameter absent in perturbative QCD, which has only a single cutoff.  
As $(\Delta\theta,\eps)\to(0,0)$ the limit is therefore direction--dependent and \emph{indefinite}.  

\paragraph{Local tangent--plane analysis.}  
Normalizing the parametric shifts to physical displacements $a,b$ in the local tangent plane at $C(\theta)$,
\begin{smalleq}
   && \Delta\theta |\dot C| = a,\\
   && \eps |N| = b,
\end{smalleq}
the estimate~\eqref{singterm} becomes
\begin{smalleq}
   && \frac{\sigma |\dot C| a}{b^2 \alpha^2 + a^2}, \qquad
   \alpha = \frac{|\dot C|}{|N|}.
\end{smalleq}
The singularity thus depends on the direction $b/a$ of approach in the tangent plane and, moreover, on the nonlocal normal vector $N_\mu$, given by the principal value of the Cauchy integral~\eqref{innerNormal} over the whole loop. This intrinsic nonlocality makes it very difficult to design a correction term to cancel the effect.

\paragraph{What a corrective term would require.}
The singularity we find in the loop equation for the minimal area is characteristic of \emph{surface} functionals, and not of the ordered loop integrals  $\iiint\prod_i df_i,\dot C_{\mu_i}(f_i)$ with tensor coefficient functions depending only on the differences $C(f_i)-C(f_j)$, as in gauge theory.
By contrast, the normal vector $N_\mu$ in \eqref{innerNormal} belongs to a different class: its coefficient is a Cauchy kernel, which depends on the parametrization function $\tau(t)$, itself defined nonlinearly in terms of the loop $C$ via \eqref{parametrization}.
This structure makes the minimal--surface singularity fundamentally nonlocal, unlike conventional Stokes--type functionals that one might try to use for cancellation.
To reproduce such a nonlocal singularity, any correction would itself have to be a genuine \emph{surface} functional, with density localized near the boundary and scaling more slowly than the area, yet carrying the same dependence on the boundary normal.
Once such functionals are admitted, however, the Hodge--dual area provides the natural resolution: its self--duality cancels these directional singularities identically.

\paragraph{Symmetry obstruction.}  
The loop equation is a \emph{vector} equation in four dimensions, whereas the ordinary area is a scalar. A scalar correction can only solve all vector components if an additional symmetry enforces it. For the self--dual surface such a mechanism exists: the tangent--area tensor has six independent components, three vanish by self--duality, and the remaining three are annihilated by the loop--space Bianchi identity. The ordinary minimal surface enjoys no such protection.

\paragraph{Renormalization--scheme dependence.}  
Renormalizability requires the string tension $\sigma$ to be scheme--independent. For the Weierstrass surface, however, estimates show strong scheme dependence: e.g.\ in dimensional regularization all perimeter divergences vanish, but the loop--equation limit remains indefinite as $\Delta\theta,\delta\to0$.

\medskip
In summary: in QCD the mismatch of singularity scaling, the two--cutoff problem, the symmetry obstruction, and scheme dependence make it implausible that subleading corrections could ``save'' the ordinary minimal surface.  
In the gradient--flow case, none of these issues arise --- there are no contact terms or short--distance singularities --- and the self--dual surface provides an exact fixed point for arbitrary loop size, while the ordinary minimal surface still fails cleanly.

\hypertarget{sec-hodge-dual-surface-as-a-fixed-point}{}
\section{ Hodge-dual surface as a fixed point}
\pdfbookmark[1]{Hodge-dual surface as a fixed point}{Hodgesurface}
\hypertarget{Hodgesurface}{}
\label{Hodgesurface}

We now turn from the general loop-calculus framework to the explicit 
construction of the Hodge-dual minimal surface and its role as an 
exact fixed point of the gradient flow for the Wilson loop.  
This part begins by formulating the embedding, boundary conditions, and 
Hodge-duality constraint, then demonstrates that the resulting surface 
satisfies the loop equation without singularities or contact terms.  
The presentation culminates with a solvable class — the planar loop, where the Hodge dual area reduces to the planar area times a universal constant.
In general case we prove some important inequalities limiting the area of the Hodge-dual surface.

\hypertarget{sub-self-dual-matrix-surface-satisfies-the-fixed-point-loop-equation}{}
\subsection{Self-dual matrix surface satisfies the fixed point loop equation}
\pdfbookmark[2]{Self-dual matrix surface satisfies the fixed point loop equation}{factorization}

Let us show that the minimal area $|S[C]|$ of the self-dual matrix surface $S$ of the disk topology satisfies the fixed point loop equation exactly, due to a geometric identity rooted in the algebraic structure of the area element. Due to the Leibniz rules \eqref{LeibnizRules1}, \eqref{LeibnizRules2} any function of the area solves the fixed point loop equation. However, only the exponential
\begin{smalleq}
    W[c,\infty] = \exp{-\kappa |S[C]|}
\end{smalleq}
satisfies the boundary conditions at infinity. Let us verify this statement.

Consider two loops far away from each other, connected with backtracking wires, like a pince-nez (see Figure \ref{fig::PinceNez}). 
\pctWPDF{0.5}{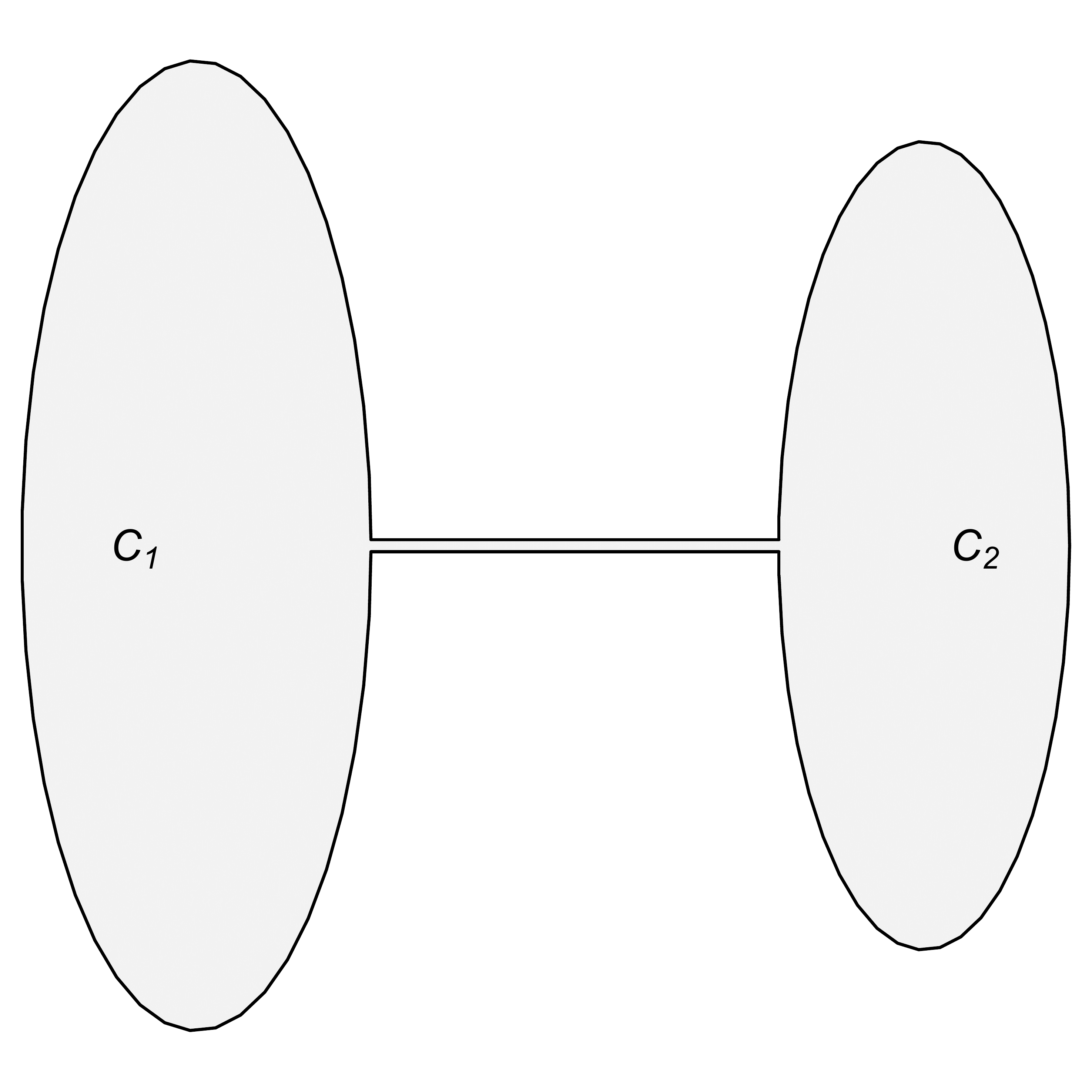}{The pair of separated loops connected with  "wires"}
This Wilson loop must factorize into a product of two
\begin{smalleq}\hypertarget{factorization}{}
\label{factorization}
    W[C_1+ \textrm{wires} + C_2] \to W[C_1]W[C_2];
\end{smalleq}
This requires (asymptotically) additive functional $|S[C]|$ in exponential, which happens for the area. It also happens for the perimeter, but perimeter is not a Stokes-type functional, so it does not satisfy the loop equation.

The main problem is to find such a functional -- as the ordinary minimal area does not satisfy the loop equation as we have seen.

\hypertarget{sub-the-definition-of-the-hodge-dual-matrix-surface}{}
\subsection{The definition of the Hodge-dual matrix surface}
\pdfbookmark[2]{The definition of the Hodge-dual matrix surface}{scalarArea}

In our loop calculus framework, the loop space diffusion operator acts as a differential constraint on any functional of the loop $C(t)$ whose variation defines a well-behaved area element. More precisely, this operator \eqref{diffusionOperator} is constructed from three functional derivatives with respect to the tangent vector $\dot C^\mu(t)$, and it measures the discontinuity of the third functional derivative along the loop. It acts on the area derivative as follows:
\begin{smalleq}
\mathcal{L}_\nu(t) |S| := \text{Disc} \left( \ff{\dot C_\mu(t)} \right) \frac{\delta |S|}{\delta \sigma_{\mu\nu}(t)}.
\end{smalleq}
The minimal surface is embedded in a higher-dimensional product space $\mathbb{R}^4 \otimes \mathbb{R}^{4}$ with real coordinates $X_\mu^A(\xi_1,\xi_2)$, where $\mu = 1,2,3,4$ labels directions in $\mathbb{R}^4$, and $A =0, 1,2,3$ labels components in the extended target space $\mathbb{R}^4$ corresponding to the algebra $\mathbb{R} \uplus \mathfrak{su}(2)$. In matrix notation, the parametric equation for this surface reads
\begin{smalleq}
    &&X_{\mu} =\tau_A X^A_\mu(\xi_1, \xi_2) = X^0_\mu(\xi_1, \xi_2) + \tau_a X^a_\mu(\xi_1, \xi_2)
\end{smalleq}  
We use the Einstein rule of implied summation over repeated indices. The area element of this surface has the form
\begin{smalleq}
    &&d \Sigma_{\mu\nu}  = d \xi_1 d\xi_2 \Sigma_{\mu\nu};\\
    &&\Sigma_{\mu\nu} = e_{l m}\pd{l} X^A_{\mu}\pd{m} X^A_{\nu};
\end{smalleq}
and the total scalar area to be minimized with Hodge-duality constraint:
\begin{smalleq}\hypertarget{scalarArea}{}
\label{scalarArea}
    &&|S| = \int_{\mathcal D} |d \Sigma| = \int_{\mathcal D} \sqrt{\oh(\Sigma_{\mu\nu})^2}  ;\\
    && \Sigma_{\mu\nu} = \oh e_{\mu\nu\alpha\beta}\Sigma_{\alpha\beta};
\end{smalleq}
Here $\mathcal D$ is some domain in a plane, topologically equivalent to a disk.
Note that this scalar area does \textbf{not} reduce to the determinant of the $2\times 2$ metric
\begin{smalleq}
    g_{l m} = \pd{l} \bar X^A_{\mu}\pd{m} X^A_{\mu}
\end{smalleq}
This difference reflects the fact that our space is not equivalent to the multidimensional Euclidean space.
Our definition of area in our space $\mathbb{R}^4 \otimes \mathbb{R}^{4}$ is a valid positive functional, invariant with respect to the diffeomorphisms of coordinates $\xi$.

There is a well-known freedom of conformal transformation of the internal coordinates
\begin{smalleq}\hypertarget{confTrans}{}
\label{confTrans}
    \xi_1 + \I \xi_2 \Rightarrow f(\xi_1 + \I \xi_2)
\end{smalleq}
which allows mapping the boundary from the unit circle to the upper semiplane or any other conformal domain. We shall use the unit disk with polar coordinates
\begin{smalleq}
  &&  \xi_1 + \I \xi_2 = r e^{\I\theta};\\
  && 0 < r < 1; 0 < \theta < 2 \pi;
\end{smalleq}
The Dirichlet boundary values for this field are linearly related to the external loop $C(\theta) \in \mathbb{R}^4$ as follows
\begin{smalleq}\hypertarget{Dirichlet}{}
\label{Dirichlet}
&& X^A_{\mu}(1, \theta) = \Lambda^A_{\mu\nu} C_\mu(\theta);\\
    &&\Lambda^0_{\mu\nu} =  \delta_{\mu\nu}; \\
    &&\Lambda^a_{\mu\nu} = \bar \eta^a_{\mu\nu}; 
\end{smalleq}
We are using the projectors $\eta^i_{\mu\nu}, \bar\eta^i_{\mu\nu}$ by 't Hooft \cite{tHooft1976}:
\begin{smalleq}\hypertarget{etas}{}
\label{etas}
       &&  \eta^i_{\mu\nu}= (\delta _{4\nu } \delta _{i\mu }-\delta _{4\mu } \delta _{i\nu }) + e_{i\mu \nu 4};\\
  &&  \bar\eta^i_{\mu\nu}= (\delta _{4\nu } \delta _{i\mu }-\delta _{4\mu } \delta _{i\nu }) - e_{i\mu\nu 4};\\
    && \oh e_{\mu\nu\lambda\rho}\eta^i_{\lambda\rho} =  \eta^i_{\mu\nu};\\
    && \oh e_{\mu\nu\lambda\rho}\bar\eta^i_{\lambda\rho} = -\bar \eta^i_{\mu\nu};
\end{smalleq}
The 't Hooft symbols $\eta^i_{\mu\nu}, \bar\eta^i_{\mu\nu} i=1,2,3$ form the basis of dual or anti-dual tensors depending on the sign in front of the $e$ tensor.
\begin{smalleq}
    &&-\tr \eta^i \cdot \eta^j = 4 \delta_{i j};\\
    &&-\tr \bar \eta^i \cdot \bar\eta^j = 4 \delta_{i j};\\
    && -\tr \bar \eta^i \cdot \eta^j = 0;
\end{smalleq}
One may verify that these matrices $\Lambda^A_{\mu\nu}$ represent orthogonal matrices for every $A =0,1,2,3$:
\begin{smalleq}
   \sum_{\mu=1}^4 \Lambda^A_{\mu\alpha} \Lambda^A_{\mu\beta} = \delta_{\alpha\beta} ;\forall A =0,1,2,3;
\end{smalleq}
The rotations of our minimal surface including the bounding loop now involve internal rotation of the $X$
\begin{smalleq}\hypertarget{O4O3}{}
\label{O4O3}
&&X^0_\mu \Rightarrow  R_{\mu\nu} X^0_\nu;\\
    &&X^a_\mu \Rightarrow O_{a b} R_{\mu\nu} X^b_\nu;\\
    &&C_\mu \Rightarrow  R_{\mu\nu} C_\nu;\\
    && O \in SO(3);\\
    && R \in SO(4);
\end{smalleq}
Note that we mapped into internal target space one of the $SU_2 $ subgroups of the rotation group $SO(4)$. 

\paragraph{Boundary condition fixing the conformal gauge.}

To fix the conformal gauge ambiguity associated with the coordinate transformation \eqref{confTrans}, we impose a boundary condition on the induced Euclidean metric at the boundary of the unit disk \( D \):
\begin{smalleq}\hypertarget{metricAtC}{}
\label{metricAtC}
    g_{l m}|_{\partial D} = \left( \pd{l} \bar X^A_\mu \, \pd{m} X^A_\mu \right)_{\partial D} \propto \delta_{l m}.
\end{smalleq}
This metric is not involved in the area functional, which is instead defined via the tensor-valued area element \(\Sigma_{\mu\nu}\). Specifically, the scalar area is given by
\[
\int_D d^2\xi\, \sqrt{\tfrac{1}{2} \left( \Sigma_{\mu\nu} \right)^2 },
\]
and does not reduce to \(\sqrt{\det\norm{g_{l m}}}\) used in the Nambo-Goto string Action \cite{Nambu1970, Goto1971}. This reflects the fact that our target space \( \mathbb{R}^4 \otimes \mathbb{R}^4 \) is not a Euclidean 16-dimensional manifold.

Nevertheless, the boundary condition \eqref{metricAtC} serves to fix the residual conformal freedom in parametrizing \( \partial D \). Under a conformal transformation \( \xi \mapsto f(\xi) \), the tensor area \( \Sigma^{A B}_{\mu\nu} \) rescales by the Jacobian \( |f'(\xi)| \), leaving the scalar area and Hodge-duality constraints invariant. However, the Dirichlet condition \eqref{Dirichlet} induces a reparametrization of the boundary loop \( C(\xi) \Ra C(f(\xi)) \), introducing a gauge ambiguity in the variational problem.

By requiring that the boundary metric be proportional to the identity, we eliminate this residual reparametrization invariance. Explicitly, we remove the ambiguity under linear transformations:
\begin{smalleq}
     &&g(\partial D) \Ra U  g(\partial D) U^{T};\\
     && U_{i l} = \pbyp{\tilde \xi_i}{\xi_l}.\\
     &&\sqrt{\det g(\partial D)} \Ra \det U \sqrt{\det g(\partial D)}
\end{smalleq}
The use of the induced Euclidean metric at the boundary as a gauge-fixing condition—despite the area functional being defined in terms of a non-Euclidean target space—also appears in generalized Plateau problems on Riemannian manifolds~\cite{Morrey1948}.

\paragraph{Mapping the physical space into the internal space}
The surface is embedded not in the Cartesian product space
$\mathbb{R}^4 \times \mathbb{R}^4$, but rather in the tensor product
$\mathbb{R}^4 \otimes \mathbb{R}^4$. This reflects the fact that
$X^A_\mu(\xi)$ is a matrix-valued vector field: for each spacetime
index $\mu = 1,\dots,4$, there are $\mathcal{N} = 4$ internal components
indexed by $A$, taking values in a representation of the $U(2)$ algebra.

Our mapping of physical space $\mathbb{R}^4$ into the internal symmetry
space $\mathbb{R}^4$ is analogous to the structure of the classical
Yang–Mills instanton, where the gauge field $A_\mu^a(x)$ maps spacetime
coordinates into the same internal space $\mathbb{R}^4$ using the 't Hooft
symbols $\eta$. Like the instanton, our minimal surface field is subject to
a local Hodge-duality constraint — a condition that will be detailed in the
next section.
In summary, the minimal Hodge-dual area represents a positive rotation-and parametric-invariant functional of the bounding loop $C$, so it could serve as a minimal area in the context of QCD.

\hypertarget{sub-the-loop-equations-and-duality-constraint}{}
\subsection{The loop equations and duality constraint}
\pdfbookmark[2]{The loop equations and duality constraint}{dualAreaFunc}

The variation of the boundary values $\delta X^A_\mu(1,\theta)$ in our area functional \eqref{scalarArea} brings the corresponding normalized area element. We follow the same steps as with the unconstrained minimal surface
\begin{smalleq}\hypertarget{dualAreaFunc}{}
\label{dualAreaFunc}
    &&|S| = \min_{\lambda, X} \int d^2\xi \left(\sqrt{\oh\tr \Sigma\cdot \Sigma^T} + \lambda_i \bar \eta^i_{\mu\nu} \Sigma_{\mu\nu}\right);\\
    && \delta |S| = \int d^2\xi \left( t_{\mu\nu} + \lambda_i \bar \eta^i_{\mu\nu}\right) 2 e_{l m} \pd{l}\delta X^A_\mu \pd{m} X^A_\nu;\\
    && t_{\mu\nu}(r,\theta) = \frac{\Sigma_{\mu\nu}}{\sqrt{2\tr \Sigma\cdot \Sigma^T}}
\end{smalleq}
By virtue of minimality conditions (which we study later) and the Stokes theorem, this variation reduces to the boundary terms
\begin{smalleq}
    &&\delta|S| = \oint_{\partial \mathcal D} d \theta \delta X^A_\mu\pd{\theta} X^A_\nu 2 T_{\mu\nu} ;\\
    && T_{\mu\nu} = (t_{\mu \nu} + \lambda_i \bar \eta^i_{\mu\nu});
\end{smalleq}
Let us convert this variation into the area derivative in $\mathbb{R}^4$ by expressing the boundary conditions for $X$ as linear functions of $ C$
\begin{smalleq}
    &&\delta|S|=  \oint d \theta \delta C_\alpha(\theta) \dot C_\beta(\theta) 2 \Lambda^A_{\mu\alpha}  \Lambda^A_{\nu\beta}T_{\mu \nu}(1,\theta);\\
    && \Lambda^0_{\nu\beta} = \delta_{\nu\beta};\\
    && \Lambda^a_{\nu\beta} = \bar \eta^a_{\nu\beta}
\end{smalleq}
Antisymmetry of the tensor $T_{\mu\nu}$ eliminates the contribution from $\Lambda^0$ and we are left with
\begin{smalleq}
    &&\delta|S| = \oint d \theta \delta C_\alpha(\theta) \dot C_\beta(\theta) 2 \bar \eta^a_{\mu\alpha}\bar \eta^a_{\nu\beta}T_{\mu \nu}(1,\theta)
\end{smalleq}
The completeness identity for the $\bar \eta^a$ tensors
\begin{smalleq}
    \bar \eta^a_{\mu\alpha}\bar \eta^a_{\nu\beta} = \delta_{\mu\nu}\delta_{\alpha\beta} -\delta_{\mu\alpha}\delta_{\nu\beta}
\end{smalleq}
simplifies the last relation to the desired form of area derivative by $C \in \mathbb{R}^4$
\begin{smalleq}
    &&\delta|S|   = \int d \theta \delta C_\alpha(\theta) \dot C_\beta(\theta)2 T_{\alpha\beta}(1,\theta);\\
    && \fbyf{|S|}{\sigma_{\alpha\beta}(\theta)} = 2 T_{\alpha\beta}(1,\theta);
\end{smalleq}
The Hodge-duality of the area derivative in \( \mathbb{R}^4 \) would require that at the boundary
\begin{smalleq}\hypertarget{Hodgeduality}{}
\label{Hodgeduality}
\bar\eta^a_{\mu\nu}\, T_{\mu\nu}(1,\theta) = 0 \quad \forall a = 1,2,3.
\end{smalleq}
As the tensor $t_{\mu\nu}$ is Hodge-dual, this condition requires the Lagrange multiplier $\lambda_i$ to vanish at the boundary
\begin{smalleq}
    \lambda_i(1,\theta) =0;
\end{smalleq}
It is a well-known fact in differential geometry that a simple bivector in four dimensions cannot be self-dual unless it vanishes. The six-dimensional space of bivectors decomposes into a direct sum of two three-dimensional subspaces of self-dual and anti-self-dual 2-forms. A simple (decomposable) bivector cannot lie entirely within the self-dual subspace unless it is zero~\cite{ATMP1999,Batista2013}. Therefore, any nonzero self-dual form must be a linear combination of multiple simple bivectors.

In particular, this means that the Hodge-duality constraint on \( t_{\mu\nu} \) cannot be satisfied by a single-valued surface field \( X_\mu(\xi) \) in \( \mathbb{R}^4 \); the field must have internal degrees of freedom. This is precisely the role of the matrix-valued embedding \( X^A_\mu(\xi) \) in our construction.

The key identity we need to prove the loop equation is that the area derivative is constructed from two dot-functional derivatives of $C^\mu(t)$ in our loop calculus:
\begin{smalleq}
&&\ff{\dot C_{\lb{\mu}}(\theta-)}  \ff{\dot C_{\rb{\nu}}(\theta+)}|S| = 2 t_{\mu\nu}(1,\theta);
\end{smalleq}
This Hodge-duality of $t_{\mu\nu}$ leads to the conclusion that our minimal area satisfies the fixed point loop equation by the Bianchi identity \eqref{ident} we have proven in section \ref{Bianchi}.
\begin{smalleq}
   &&\pd{\mu} \fbyf{|S|}{\sigma_{\mu\nu}(\theta)} 
   \propto e_{\mu\nu\lambda\rho}\left(\ff{\dot C_\mu(\theta-)}-\ff{\dot C_\mu(\theta+)}\right)\nonumber\\
   &&\ff{\dot C_{\lambda}(\theta-)}  \ff{\dot C_{\rho}(\theta+)} |S|\equiv 0
\end{smalleq}
In our formalism, this is not merely a geometric analogy — it is an exact operator identity in loop space. The Bianchi identity holds because the area derivative is explicitly built from dot-functional derivatives, which obey algebraic identities that mirror differential geometry.

This is the central advantage of the self-dual surface: it is constructed so that the variation of its area with respect to loop deformations always satisfies the Bianchi identity. As a result, the loop operator annihilates the corresponding Wilson functional:
\begin{smalleq}
&&\mathcal{L}_\nu(t) \, \exp{-\kappa S[C]} =  \pd{\mu} \left(\fbyf{S[C]}{\sigma_{\mu\nu}}\exp{-\kappa S[C]}\right) =\nonumber\\
   &&-\kappa\exp{-\kappa S[C]} \pd{\mu}\fbyf{S[C]}{\sigma_{\mu\nu}} =0
\end{smalleq}
In the last equation we used the second Leibniz rule and the fact proven in loop calculus (section \ref{LoopCalculus}) that $\pd{\mu}F[C] =0$ for any Stokes functional $F[C]$ including $\exp{-\kappa S[C]}$.
This confirms that the self-dual surface is an exact classical solution of the loop equation, free of singularities, contact terms, or ambiguities.
\paragraph{Sufficiency, not necessity.}
The fixed–point loop equation requires only 
\(\partial_\mu\!\left(\delta W/\delta\sigma_{\mu\nu}\right)=0\).
Hodge self–duality of the area derivative is a \emph{sufficient} mechanism:
by the loop–space Bianchi identity (Sec.~\ref{Bianchi}) it makes the left-hand
side vanish identically. It is not a \emph{necessary} condition; other, non
self-dual solutions might also satisfy the divergence constraint. Our
construction uses self-duality to \emph{guarantee} the loop equation. By
contrast, the conventional minimal (Nambu–Goto/Douglas) area fails the
smooth loop-equation test at the final stage, when the operator $\partial_\mu(\theta)$ is applied to its area derivative $t_{\mu\nu}(1,\theta)$ (see the earlier subsection on
its violation).

In the next section, we study this Hodge-dual minimal surface in some detail, and present a solvable example with a circular bounding loop $C$.
\hypertarget{sub-minimal-surface-equations}{}
\subsection{Minimal Surface Equations}
\pdfbookmark[2]{Minimal Surface Equations}{MinSurfEq}
\hypertarget{MinSurfEq}{}
\label{MinSurfEq}

To derive the Euler–Lagrange equations, we treat \( \partial_r X^A_\mu \) and \( \partial_\theta X^A_\mu \) as independent variables and introduce Lagrange multipliers to impose the self-duality and zero-curl constraints. The variational principle is
\begin{smalleq}\hypertarget{HodgeArea}{}
\label{HodgeArea}
|S| = \int_0^1 dr \oint d\theta \left[
    \sqrt{\oh(\Sigma_{\mu\nu})^2}
    + \lambda_i \bar{\eta}^i_{\mu\nu} \Sigma_{\mu\nu}
    + \rho^A_\mu e_{lm} \partial_l X^A_{\mu,m}
\right],
\end{smalleq}
with
\begin{smalleq}
   && \Sigma_{\mu\nu} = e_{lm} X^A_{\mu,l}X^A_{\nu,m}.
\end{smalleq}

The variation yields the equation for \( X^A_{\mu,m} \). In matrix notation:
\begin{smalleq}
   && \left(t + \lambda_i \bar{\eta}^i\right)\cdot X^A_{m} = \partial_m \rho^A,\\
    && t = \frac{\Sigma}{\sqrt{2\tr \Sigma\cdot \Sigma^T}},
\end{smalleq}
along with the self-duality constraint (equation for $\lambda_i$)
\begin{smalleq}\hypertarget{sigmaDual}{}
\label{sigmaDual}
    \tr \bar{\eta}^i \cdot  \Sigma = 0,
\end{smalleq}
and the curl-free condition (equation for $\rho^A_\mu$). 
\begin{smalleq}\hypertarget{rhoeq}{}
\label{rhoeq}
     \pd{l} \left(t + \lambda_i \bar{\eta}^i \right)^{-1} e_{lm}\pd{m} \rho^A= 0.
\end{smalleq}
Note  that inversion of the antisymmetric $4\times 4$ here is a simple algebraic operation using Pfaffian
\begin{smalleq}
    \left(t_i \eta^i + \lambda_i \bar{\eta}^i \right)^{-1} = \frac{t_i \eta^i - \lambda_i \bar{\eta}^i}{ \lambda_i^2 -  t_i^2}
\end{smalleq}
We adopt the following Dirichlet boundary conditions:
\begin{subequations}\hypertarget{BC}{}
\label{BC}
\begin{smalleq}
    &&X^A_\mu(1,\theta) = \Lambda^A_{\mu\nu} C_\nu(\theta)\;;\\
     && \Lambda^0_{\nu\beta} = \delta_{\nu\beta};\\
    && \Lambda^a_{\nu\beta} = \bar \eta^a_{\nu\beta};
\end{smalleq}
\end{subequations}
together with boundary condition on the Euclidean metric tensor
\begin{smalleq}\hypertarget{indicedmetric}{}
\label{indicedmetric}
    && g_{l m}(r,\theta) = X^A_{\mu,l}X^A_{\mu,m};\\
    \hypertarget{gcond}{}
\label{gcond}
    && g_{l m}(1,\theta) \propto \delta_{l m}
\end{smalleq}
Finally, there is a boundary condition for the Lagrange multiplier:
\begin{smalleq}\hypertarget{lambdacond}{}
\label{lambdacond}
    \lambda_i(1,\theta) =0;
\end{smalleq}
\hypertarget{sub-scale-invariance-and-area-law}{}
\subsection{Scale Invariance and Area Law}
\pdfbookmark[2]{Scale Invariance and Area Law}{scalingLaw}

Although we lack a general solution to this nonlinear system, it obeys an important scaling property.

All the constraints (Hodge-dualty, conformal metric at the boundary) are homogeneous-- they do not change when the coordinates of the surface are rescaled. The area itself is a second order homogeneous functional of the coordinates, and the boundary conditions are linear relations between $X$ and the loop $C$.  

This scaling property makes the minimal Hodge-dual area a second order homogeneous nonlinear functional of the bounding loop $C$. This functional, unlike the Douglas functional for the Euclidean minimal surface \cite{Douglas1931,Osserman2002}, is not a quadratic form, but still, it scales as a square of the size of the loop.

Let the boundary loop \( C \) be rescaled as
\begin{smalleq}
    C_\mu(\theta) = \sqrt{A_{\min}[C]} \, c_\mu(\theta),
\end{smalleq}
where \( A_{\min}[C] \) is the Euclidean minimal area enclosed by \( C \). The normalized loop \( c_\mu(\theta) \) captures the shape, independent of scale.

Rescaling the surface accordingly:
\begin{smalleq}
X^A_\mu(\xi) = \sqrt{A_{\min}[C]} \, x^A_\mu(\xi), \quad
x^A_\mu(1,\theta) = \Lambda^A_{\mu\nu} c_\nu(\theta),
\end{smalleq}
the area becomes
\begin{smalleq}
|S| = A_{\min}[C]  \min \int d^2\xi \sqrt{\oh (e_{lm} \partial_l x^A_{\mu} \partial_m x^A_{\nu})^2}.
\end{smalleq}

Since the equations for the rescaled variables are scale-invariant, we conclude:
\begin{smalleq}\hypertarget{scalingLaw}{}
\label{scalingLaw}
    |S| = A_{\min}[C]  s[c],
\end{smalleq}
where \( s[c] \) depends only on the shape of the loop.

In the next section, we construct an exact solution for an circular planar loop with fixed \( A_{\min}[C] \). The resulting ratio turns out to be a universal constant in this case.

Naturally, \( s[c] \) \textbf{must} in general depend on the loop shape. If it were constant, the area functional would be proportional to the Euclidean minimal area, implying proportional area derivatives. But this is impossible: the self-dual surface yields a self-dual area derivative, whereas the Euclidean minimal surface includes an anti-self-dual component. Therefore, \( s[c] \) cannot be a constant functional of \( C \).

\hypertarget{sub-minimal-surface-bounded-by-a-circular-loop}{}
\subsection{Minimal surface bounded by a circular loop}
\pdfbookmark[2]{Minimal surface bounded by a circular loop}{circular}
\hypertarget{circular}{}
\label{circular}

There is a special solution to the general problem of self-dual minimal surfaces.
This solution corresponds to all parameters $\lambda_i$, $t_{\mu\nu}$ being constant across the surface. In that case, the equation~\eqref{rhoeq} is identically satisfied, without restricting $\rho^A_\mu$. The vector $\lambda_i$ also remains arbitrary, so we can set it to zero to satisfy the boundary condition \eqref{lambdacond}.

We treat the shape of the boundary curve $C(\theta)$ as part of the solution. Among all possible boundary shapes, the circular loop provides the simplest setting where all calculations can be carried out explicitly, while still satisfying the full self-duality and loop constraints. We take:
\begin{smalleq}
&&C_\nu(\theta) = \left\{\cos\theta, \sin\theta, 0, 0\right\}, \\
&&A_{\mathrm{min}}[C] = \pi
\end{smalleq}
Let us switch to matrix notations.
We assume the following general ansatz:
\begin{smalleq}
    &&X^A(r,\theta) = V^A(r) \cdot C(\theta);
\end{smalleq}
with some tensor $V^A(r)$.
Then the area element becomes:
\begin{smalleq}
  \Sigma_{\mu\nu} = e_{l m}\pd{l}X^A_{\mu} \pd{m}X^A_{\nu} = \pd{r} V^A \cdot \mathcal E \cdot V^A -  V^A \cdot \mathcal E \cdot \pd{r}V^A 
\end{smalleq}
Here $\mathcal E$ is the antisymmetric 4-tensor with components \(   \mathcal E_{12} = -  \mathcal E_{21} = 1 \), and all other components zero.
This tensor comes from the circle equation
\begin{smalleq}
    C_{\lb{\alpha}}(\theta)\dot C_{\rb{\beta}}(\theta) = \mathcal E_{\alpha\beta}
\end{smalleq}
We are looking for a solution linear in \( r \), satisfying our boundary condition:
\begin{smalleq}
&&V^A(r) = \Lambda^A  + (r - 1) v^A ;
\end{smalleq}

This produces two vector equations (one for each of the two terms of expansion in $r$). 
\begin{smalleq}\hypertarget{circleDuality}{}
\label{circleDuality}
   && \tr \bar \eta^k \cdot \left(\pd{r} V^A \cdot \mathcal E \cdot V^A -  V^A \cdot \mathcal E \cdot \pd{r}V^A\right) =0;\nonumber\\
   && \forall k=1,2,3, r = 0,1
\end{smalleq}
We solve these equations and compute the area $S[C]$ in the \Mathematica{}  notebook \cite{DualityEquations}.

The six equations \eqref{circleDuality} and the boundary conditions are all satisfied by
\begin{smalleq}
   && V^0_{\mu\nu}(r) = r \delta_{\mu\nu};\\
   && V^a_{\mu\nu}(r) = r \bar \eta^a_{\mu\nu};
\end{smalleq}
There are two solutions of all these equations for the set of three parameters in the Anzatz for $v^A$, we have chosen the solution with smaller area.
The induced Euclidean metric tensor reduces to the following 
\begin{smalleq}
  g(r) = \left(
\begin{array}{cc}
 4 & 0 \\
 0 & 4 r^2 \\
\end{array}
\right)
\end{smalleq}
The tensor $\Sigma_{\mu\nu}$ comes out as follows
\begin{smalleq}
    &&\Sigma =2r\left(
\begin{array}{cccc}
 0 & 1 & 0 & 0 \\
 -1 & 0 & 0 & 0 \\
 0 & 0 & 0 & 1 \\
 0 & 0 & -1 & 0 \\
\end{array}
\right);\\
&& t = \frac{1}{\sqrt{2}}\left(
\begin{array}{cccc}
 0 & 1 & 0 & 0 \\
 -1 & 0 & 0 & 0 \\
 0 & 0 & 0 & 1 \\
 0 & 0 & -1 & 0 \\
\end{array}
\right);
\end{smalleq}
The minimal area element vanishes at $r=0$ which eliminates the singularity at the origin.
The total area, after integration over $r,\theta$ reads
\begin{smalleq}
    |S| = \oint d \theta \int_0^1 d r \sqrt{\oh \tr \Sigma\cdot \Sigma^T} = 4\sqrt{2} \pi \int_0^1 r d r = 2 \sqrt{2} \pi
\end{smalleq}
This makes the coefficient
\begin{smalleq}
    s[c]_{\mathrm{circle}} = 2 \sqrt{2}
\end{smalleq}

\hypertarget{sub-minimal-surface-bounded-by-an-arbitrary-planar-loop}{}
\subsection{Minimal surface bounded by an arbitrary planar loop}
\pdfbookmark[2]{Minimal surface bounded by an arbitrary planar loop}{exactPlanar}
\hypertarget{exactPlanar}{}
\label{exactPlanar}
The above solution for the Hodge-dual minimal surface bounded by a circle can be generalized to an arbitrary non-intersecting planar loop.

\theorem{The minimal dual area for any non-intersecting planar loop equals $2\sqrt{2}$ times the planar area inside this loop.}

\begin{proof}
Let us use the conformal invariance of the dual area functional \eqref{dualAreaFunc} and map the unit circle onto the given planar bounding loop $\mathbb S_1 \mapsto C$. 
After that, the minimal surface can use the coordinates $\xi = (x_1,x_2) $ inside that planar loop $C$ as internal coordinates of the surface.
In the context of our surface embedded into $\mathbb{R}^4 \otimes \mathbb{R}^4$, this parametrization reduces to
\begin{smalleq}
   &&X^A_{l} = \Lambda^A_{l m} \,\xi_m, \quad l,m = 1,2,\\
   &&X^A_{\dot l} = Y^A_{\dot l}(\xi), \quad \dot l = 3,4,\\
   &&\xi \in \mathcal{D}, \quad C = \partial \mathcal{D},
\end{smalleq}
where $\Lambda^A_{\mu\nu}$ are defined in \eqref{BC}, and the new field $Y^A_{\dot l}(\xi)$ satisfies trivial boundary conditions:
\begin{smalleq}\hypertarget{YBC}{}
\label{YBC}
    Y^A_{\dot l}(\xi \in C) = 0.
\end{smalleq}

The area functional is
\begin{smalleq}
   && |S| = \int_{\mathcal{D}} d^2\xi \sqrt{\tfrac12 \Sigma_{l m}^2 + \Sigma_{l \dot l}^2 + \tfrac12 \Sigma_{\dot l \dot m}^2},\\
   && \Sigma_{l m} = e_{a b} \Lambda^A_{a l}\Lambda^A_{b m},\\
   && \Sigma_{l \dot l} = e_{a b}\Lambda^A_{a l} \,\partial_{b} Y^A_{\dot l},\\
   && \Sigma_{\dot l \dot m} = e_{a b}\,\partial_{a} Y^A_{\dot l} \,\partial_{b} Y^A_{\dot m}.
\end{smalleq}
With zero boundary values for the $Y$ field, this area functional attains its absolute minimum at $Y^A_{\dot l}(\xi) = 0$. The non-negative term $\Sigma_{l \dot l}^2 + \tfrac12 \Sigma_{\dot l \dot m}^2$ vanishes only for a constant $Y^A(\xi)$, but the boundary conditions \eqref{YBC} in that case restrict this constant tensor $Y^A(\xi)$ to zero.

It remains to verify that the Hodge-duality condition is satisfied:
\begin{smalleq}
    \sum_{l, m =1,2} \eta^k_{l m} \,\Sigma_{l m} = 0, \quad \forall\, k = 1,2,3.
\end{smalleq}
We verify this identity in a \Mathematica{} notebook \cite{DualityEquations}; it is the same identity that ensures self-duality for the circular solution. One can readily check that this general planar solution reduces to this circular solution for a disk where $\xi = r\{\cos\theta,\sin\theta\}$.

The normalization of the area element is the same as in the circular case:
\begin{smalleq}
    &&\sqrt{\tfrac12 \Sigma_{l m}^2} = 2\sqrt{2},\\
    &&\left|S\left[C_{\mathrm{planar}}\right]\right| = 2\sqrt{2} \,|\mathcal{D}|.
\end{smalleq}
\end{proof}\par\normalfont\normalsize

\hypertarget{sub-basic-inequalities-needed-for-quark-confinement}{}
\subsection{Basic inequalities needed for quark confinement}
\pdfbookmark[2]{Basic inequalities needed for quark confinement}{parityBreaking}

We prove in this section that, for an arbitrary loop $C$, the self-dual surface area is bounded by some calculable minimal area in the linear space. 

Here is the lower bound:
\begin{theorem}
The area of the constrained matrix surface (i.e., with the duality constraint) is not smaller than that of the minimal unconstrained matrix surface with the same bounding curve $X^A(1,\theta) = \Lambda^A \cdot C(\theta)$.
\end{theorem}

\begin{proof}
The duality constraint restricts the allowed configurations $X^A_\mu(\xi)$ to those for which the  local area element satisfies the self-duality condition \eqref{Hodgeduality} at every point of the surface.

In contrast, the unconstrained configuration space includes arbitrary area element $\Sigma_{\mu\nu}$ with no local duality structure. The self-dual (or anti-self-dual) sector therefore forms a proper subset of the unconstrained space.

Since the constrained space is strictly smaller, the infimum of the area functional over that space cannot be less than the infimum over the unconstrained space:
\begin{smalleq}
   \inf_{X \in \mathbb{R}^4 \otimes \mathbb{R}^{4},\, \Sigma = \pm *\Sigma} |S| \ge \inf_{X \in \mathbb{R}^4 \otimes \mathbb{R}^{4}} |S|.
\end{smalleq}
\end{proof}\par\normalfont\normalsize
The unconstrained minimal area in $\mathbb{R}^4 \otimes \mathbb{R}^{4}$ can  also be bounded.
\begin{theorem}
The area of the unconstrained matrix surface (i.e., without the duality constraint) is positive but not larger than four times the minimal surface area in Euclidean space with the bounding curve $C \subset \mathbb{R}^4$.
\end{theorem}

\begin{proof}
The scalar area element $ \abs{\Sigma_{\mu\nu}}$  satisfies the inequality:
\begin{smalleq}
 && \abs{\Sigma_{\mu\nu}} = \sqrt{\oh\sum_{\mu,\nu} \left(\sum_A e_{l m}\pd{l}X^A_\mu \pd{m}X^A_\nu\right)^2 } \le \nonumber\\
 &&\sum_{A} \sqrt{\oh\sum_{\mu,\nu} \left( e_{l m}\pd{l}X^A_\mu \pd{m}X^A_\nu\right)^2 } = \sum_{A}\sqrt{\det \norm{g^A_{l m}}} ;\\
  &&  g^A_{l m} = \partial_l X^A_\mu \partial_m X^A_\mu
\end{smalleq}
(In the last formulas the repeated indices $A$ were not implicitly summed over).
We conclude, that the minimum  of the unconstrained surface area cannot exceed that of  an additive functional
\begin{smalleq}
   && |\tilde S| = \int d^2 \xi \sum_{A=0}^3\sqrt{\det \norm{g^A_{l m}}};
\end{smalleq}
Now, each of four terms for $A =0,1,2,3$ in this additive functional depends of the particular component of the field $X^A_\mu(\xi) \in \mathbb{R}^4$, and it is minimized under boundary conditions $X^A_\mu(1,\theta) =\Lambda^A_{\mu\nu} C_\nu(\theta)$ for this particular component. Each of these boundary conditions corresponds to a different rotation of the same loop $C \in \mathbb{R}^4$.  This functional is minimized by the sum of four Euclidean minimal areas of rotated loops  in $\mathbb{R}^4$. The boundary condition \eqref{gcond} on the induced Euclidean metric will be automatically fulfilled by conformal metric $g^A_{l m}$ minimizing the area in each of the four $\mathbb{R}^4$. The values of each minimal area are invariant with respect to rotation and are equal to the minimal Euclidean area $A_4[C]$.
On the other hand,  the unconstrained area is positive definite and cannot be zero without breaking the Dirichlet boundary conditions. Therefore,
\begin{smalleq}
   0 < \min_{X \in \mathbb{R}^4 \times \mathbb{R}^{4}} |S| \le 4\min_{X \in \mathbb{R}^4} \int d^2\xi \sqrt{\det{\norm{\pd{l}X\cdot \pd{m} X}}}= 4 A_4[C].
\end{smalleq}
\end{proof}\par\normalfont\normalsize

The significance of the lower bound  in the first theorem is that the unconstrained problem is the conventional minimal surface in the linear space $\mathbb{R}^4 \otimes \mathbb{R}^{4}$, so it can, in principle, be solved by harmonic maps.

The upper bound for that unconstrained surface  in the second theorem is related to a well-known Euclidean minimal surface in four dimensions, solved by minimizing the Douglas functional \cite{Osserman2002}.
The second inequality is helpful in estimating the lower bound for the Hodge-dual vector matrix surface we really need to minimize. However, this \textbf{upper limit} of the lower bound  does not directly limit the Hodge-dual vector matrix area: it can be either above or below $ 4 A_4[C]$. In particular, for a planar loop with topology of a circle, as we have seen, it is $ 2 \sqrt{2} A_4[C] \approx 2.82843 A_4[C]$, which is below that upper limit $4 A_4[C]$ for the unconstrained surface.

\hypertarget{sub-the-parity-breaking-problem-and-its-solution}{}
\subsection{The parity breaking problem and its solution}
\hypertarget{parityBreaking}{}
\label{parityBreaking}

The Hodge duality involves an arbitrary parity parameter $\lambda$ (Hodge chirality):
\begin{smalleq}
   * \Sigma = \lambda \Sigma, \quad \lambda = \pm 1.
\end{smalleq}
In our theory (see Section~\ref{MinSurfEq}), flipping the sign of the Hodge chirality corresponds to changing the sign of the Levi-Civita symbol $e_{\mu\nu\alpha\beta}$ in all equations.  
This makes the Hodge chirality an odd variable under a space parity transformation. The parity transformation also switches $\eta^i_{\mu\nu} \leftrightarrow \bar{\eta}^i_{\mu\nu}$ in the boundary conditions, including those used in the equation for the loop $C$.

For both signs of Hodge chirality, our boundary conditions~\eqref{BC} represent the component $X^A(1,\theta)$ as the loop $C$ rotated by a certain orthogonal matrix: for $A=0$, the identity matrix; for $A>0$, either $\bar{\eta}^A$ or $\eta^A$, depending on the Hodge chirality of the surface.

If the bounding loop explicitly depends on the Levi-Civita symbol—thereby breaking chirality symmetry—we must flip the sign of these symbols as well, e.g.,
\begin{smalleq}
   C^{\pm}_\mu(\theta) \;=\; P_{\mu}(\theta) \pm e_{\mu\nu\alpha\beta} Q_{\nu}(\theta) R_{\alpha}(\theta) S_{\beta}(\theta),
\end{smalleq}
leading to an obvious change in shape.

There are two minimal surfaces $S^{\pm}[C]$, one for each sign of $\lambda$. We can prove only that
\begin{smalleq}
    |S^{+}[C^+]| = |S^{-}[C^-]|.
\end{smalleq}
The stronger statements $|S^{+}[C^+]| \stackrel{?}{=} |S^{+}[C^-]|$ or $|S^{-}[C^+]| \stackrel{?}{=} |S^{-}[C^-]|$ do not, in general, follow from our equations.

Thus, not only does the geometric shape of $S^{\pm}[C]$ depend on its Hodge chirality, but the minimal area $|S^{\pm}[C]|$ as a functional of the bounding loop $C$ can also depend on the loop’s chirality relative to the chirality of the surface. This implies a possible parity violation in QCD, contrary to both experimental evidence and theoretical arguments.

I am indebted to Ed Witten for pointing out this issue.

To avoid the problem, we cannot simply retain a single exponential term with a fixed Hodge chirality; a symmetric combination is required. A naive sum of two exponentials would violate the factorization property~\eqref{factorization}. The only way to restore parity \emph{and} preserve factorization of the solution to the fixed point loop equation is to use the exponential of the sum:
\begin{smalleq}
    W[C] =\exp{- \kappa \, (\,|S^{+}[C]| + |S^{-}[C]|\,)}.
\end{smalleq}
This form is parity-even and factorizes at separated sub-loops. It solves the classical loop equation due to two key properties of the loop operator $\partial_{\mu} \frac{\delta}{\delta \sigma_{\mu\nu}(\theta)}$: the Leibniz rules~\eqref{LeibnizRules1}, \eqref{LeibnizRules2} and the Bianchi identity~\eqref{Bianchi}.  
In each term $|S^{\pm}[C]|$, applying the loop operator and using Hodge duality replaces it by $\pm 1$ times the Bianchi identity, yielding zero.

This exponential of the sum of two minimal areas asymptotically approaches the Nambu–Goto form $\exp{- \sigma_{\mathrm{eff}}[C]\,A_4[C]}$ with effective string tension $\sigma_{\mathrm{eff}}[C] = \kappa\,(s^{+}[C] + s^{-}[C])$, a parity-even functional of the loop shape.

\medskip
\noindent\textbf{Conclusion.}  
We have found a parity-preserving version of the Wilson loop area law that solves the fixed-point loop equation. This is an exact analytic solution for a fixed point of the loop space diffusion corresponding to spontaneous stochastization of the  \YM{} gradient flow.

\hypertarget{sec-momentum-loop-equation-and-decaying-gradient-flow}{}
\section{Momentum loop equation and decaying gradient flow}
\pdfbookmark[1]{Momentum loop equation and decaying gradient flow}{Anzatz}

Now let us come back to the nonsingular loop equation \eqref{LoopEq} and compare it with the loop equation in the \NS{} hydrodynamics \cite{M93, M23PR}. It is a four-dimensional version of the \NS{} loop equation, except the advection term of the \NS{} equation is missing here.

 The diffusion in fluid dynamics leads to a universal operator in loop space, applicable to both Abelian and non-Abelian theories.

\hypertarget{sub-the-plane-wave-in-loop-space}{}
\subsection{The plane wave in loop space }
This remarkable analogy allows us to immediately write down an Ansatz for a solution in the form of the momentum loop equation (MLE)
\begin{smalleq}\hypertarget{Anzatz}{}
\label{Anzatz}
    && W[C,\tau] = \VEV{\exp{\I \int_0^{2\pi}  d \theta \dot C_\mu(\theta) P_\mu(\theta,\tau)}}_{P(\tau)};\\
    \hypertarget{MLE}{}
\label{MLE}
    && \pd{\tau} P_\nu(\theta,\tau) = \alpha \Delta P_\mu(\theta)\bar P_{\lb{\mu}}(\theta,\tau) \Delta P_{\rb{\nu}}(\theta);\\
    && \bar P_\mu(\theta) =  \frac{P_\mu(\theta+)+P_\mu(\theta-)}{2}; \\
    && \Delta P_\nu(\theta) =  P_\nu(\theta+)-P_\nu(\theta-); 
\end{smalleq}
The brackets $\VEV{}_{P(\tau)}$ correspond to the averaging over an ensemble of solutions of the time evolution of $P_\mu(.,\tau)$ described by the above equation.

This Anzatz \eqref{Anzatz} is the loop space version of the plane wave. The loop equation \eqref{LoopEq} involving only dot functional derivatives, such an Anzatz would exactly satisfy the loop equation, with each dot derivative $\ff{\dot C_\nu(\theta)}$ equivalent to multiplication of $\I P(\theta)$.

The relation between the loop operator and the discontinuity of the momentum loop $\Delta P_\nu(\theta)$ was discovered and investigated in earlier papers in QCD \cite{MLDMig86, SecQuanM95, Mig98Hidden}. The Momentum loop equation for QCD involves the contact terms, which complicate the equation.

In case of the gradient flow, there are no contact terms, and the equation becomes algebraic (though still singular).

\hypertarget{sub-the-nature-of-the-ym}{}
\subsection{The nature of the \YM{}
\pdfbookmark[2]{The nature of the}{fsol}
 theory is hidden in initial data}
This universal MLE equation \eqref{MLE} looks too simple. Where is the gauge theory? 

The complexity of the theory and the specifics of the Lie algebra are hidden in the initial conditions $P_\mu(\theta,0)$.
The distribution of $P_\mu(\theta,0)$ is given by the path integral (functional Fourier transform)
\begin{smalleq}
    W[P,0] = \int [\delta C] \exp{-\I \oint d\theta \dot C_\mu(\theta) P_\mu(\theta)} W[C,0]
\end{smalleq}
This integral is discussed in detail in previous work and summarized in \cite{M23PR}; therefore, we do not need to revisit it here.

There are two kinds of initial conditions, which make sense.
\begin{enumerate}
\item The full \YM{} action in the initial Gibbs distribution for the gauge field. In this case, the initial data is the entire gluodynamics, largely unknown, except in Monte Carlo simulations on the lattice.
    \item The uniform gauge field $A_\mu(x) = \const{}$,with nontrivial field strength $F_{\mu\nu} = [A_\mu, A_\nu] \neq 0$; the so-called "master field". The computations of the Fourier transform were discussed in \cite{Mig98Hidden}; we shall use these methods later in this paper.
\end{enumerate}

The initial distribution of momentum in both cases depends on the invariants of the specific gauge group.
\hypertarget{sub-summary-of-the-new-loop-calculus-and-momentum-loop-equation}{}
\subsection*{Summary of the new loop calculus and momentum loop equation}
\begin{itemize}
    \item The loop calculus can be recast using only the dot derivatives $\ff{\dot C_\mu(s)}$
    \item These dot derivatives produce finite parametric invariant results when applied to the Wilson loop, with different left and right limits.
    \item The field strength is created by a wedge product of the left and right dot derivatives, and its covariant derivative -- by the difference of the left and right dot derivatives.
    \item The evolution of the Wilson loop in the gradient flow is described by the loop equation, which is trilinear in these dot derivatives.
    \item Therefore, the momentum loop Anzats \eqref{Anzatz} satisfies the algebraic ODE \eqref{MLE} with all specifics of the gauge theory hidden in the initial data.
    \item This is a dramatic reduction of dimensionality of the gradient flow, opening the door to exact solutions.
\end{itemize}
\hypertarget{sec-decaying-turbulence-as-a-fixed-trajectory-of-the-gradient-flow}{}
\section{Decaying turbulence as a fixed trajectory of the gradient flow}
\hypertarget{sub-the-decaying-solution-and-its-time-dependence}{}
\subsection{The decaying solution and its time dependence}
The gradient flows dissipate the energy functional \eqref{Energy}
\begin{smalleq}
    \pd{\tau} E = \alpha \int d^4 x \tr [D_\mu ,F_{\mu \nu}]^2 \leq 0
\end{smalleq}
This functional is negative for all fields except those satisfying the
classical field equations
\begin{smalleq}
    [D_\mu ,F_{\mu \nu}] =0
\end{smalleq}
Therefore, we can expect that there must exist a gradient flow converging to every classical solution. This is proven as Theorem 4.1 in the mathematical paper \cite{nagasawa1989asymptotic}.

The properties of the flow were not studied in that paper.
This asymptotic decay process leading to vanishing field strength is described in our theory by  a family of exact solutions of MLE
\begin{smalleq}
   && P_\mu(\theta,\tau) = \frac{f_\mu(\theta)}{\sqrt{\alpha (\tau + \tau_0)}};\\
   && \bar f_\nu \left(\Delta f_{\mu}^2-1\right) = (\Delta f_\mu   \bar f_{\mu} )  \Delta f_{\nu} 
\end{smalleq}
\hypertarget{sub-recurrent-equation-on-a-circle}{}
\subsection{Recurrent equation on a circle}
The last equation for $f_\mu(\theta)$ can be solved exactly, by a certain limiting procedure.
First, we observe that the left and the right sides are vectors times some scalar coefficients. Unless both of these coefficients vanish, these vectors are collinear. 
But in that case, the area derivative of the Wilson loop identically vanishes at every point on an arbitrary loop $C$. This implies that this is a trivial solution where the gauge field is a pure gauge.

The nontrivial solutions, with finite field strength along the decay trajectory, all correspond to both scalar coefficients vanishing for every angle $\theta$. We can rewrite these equations as
\begin{smalleq}
    && (f_\mu(\theta+) + f_\mu(\theta-))(f_\mu(\theta+) - f_\mu(\theta-))=0;\\
    && (f_\mu(\theta+) - f_\mu(\theta-))^2 =1;\\
\end{smalleq}
The first equation can be rewritten as the continuity of the length of  the vector $f_\mu(\theta)$
\begin{smalleq}
    f_\mu(\theta+)^2 = f_\mu(\theta-)^2 
\end{smalleq}
We conclude that these vectors are located on a sphere:
\begin{smalleq}\hypertarget{fsol}{}
\label{fsol}
    f_\mu(\theta) = R n_\mu(\theta); \textit{ where } n_\mu(\theta) \in \mathbb S_3
\end{smalleq}
The second equation relates the radius of the sphere to the angle between the consecutive vectors
\begin{smalleq}
    1-n_\mu(\theta-) n_\mu(\theta+)=  \frac{1}{2 R^2 } = \const{}
\end{smalleq}
There is also an important requirement of periodicity 
\begin{smalleq}\hypertarget{periodicity}{}
\label{periodicity}
    n_\mu(\theta+ 2 \pi) = n_\mu(\theta)
\end{smalleq}
\hypertarget{sub-random-walk-on-su-2}{}
\subsection{Random walk on $SU_2 $}
\pdfbookmark[2]{Random walk on}{eqbeta}

Let us start from a point $n(0) \in \mathbb S_3$ and keep applying this recurrent equation $n(\theta) \Ra n(\theta+d \theta)$, assuming each time that the next vector $n(\theta)$ is obtained from the previous one by a rotation at the same angle $\pm\beta$ in some plane. 

Such a sequence of rotations can be implemented using the quaternions. In physics notations, with Pauli matrices $\hat \tau = \left\{ \tau_1, \tau_2, \tau_3\right\}$ it is a $SU_2$ matrix associated with a unit vector $n \in \mathbb S_3$
\begin{smalleq}
   && \hat n = n_4 I + \I \hat \tau \cdot \vec n;\\
   && \oh \tr \hat n = n_4;\\
   && \hat n \hat n^\dag = \hat n^\dag \hat n =I;\\
   && \det \hat n =1;\\
   && \oh \tr (\hat n_1 \hat n^\dag_2) = \vec n_1\cdot \vec n_2
\end{smalleq}
In this representation, the relation between the vector $\vec n(\theta+d\theta)$ and the previous vector is
\begin{smalleq}
   && \hat n(\theta+) = \left(\cos\beta I + \I \sin\beta \hat \tau \cdot \vec r \right)\hat n(\theta-);\\
   && \vec r^2 = 1;\\
   && \oh \tr (\hat n(\theta+) \hat n^\dag(\theta-)) = \cos\beta
\end{smalleq}
The init vector $\vec r \in \mathbb S_2$ parametrizes such a rotation as an element of $SU_2 $.

In general, we get a random walk of angular steps $\pm\beta$ of equal geodesic length $\beta$ in a random direction on a sphere $\mathbb S_3$.

The radius of the sphere in the solution \eqref{fsol} for $f_\mu$
\begin{smalleq}
    R = \frac{1}{2 \sin\frac{\beta}{2}}
\end{smalleq}
We represented this random walk on a sphere $\mathbb S_3$ as a random walk on $SU_2 $.
\begin{smalleq}
   && \hat n(\theta) = \prod_{l=1}^{k-1} \left(\cos(\beta) I + \I \sin(\beta) \hat \tau \cdot \vec r_l \right); \; \forall{ \theta_k < \theta < \theta_{k+1}}
\end{smalleq}
\hypertarget{sub-the-discontinuities}{}
\subsection{The discontinuities}
This inequality $\theta_k < \theta < \theta_{k+1}$ means that the matrix $\hat n(\theta)$  is piecewise constant. It gets a new matrix factor in the product at a value $\theta_k$, and stays constant until the next discrete value $\theta_{k+1}$.

As we have seen above, these discontinuities are essential for the existence of the area derivative of the Wilson loop, which represents the field's strength.

In the end, we should tend to infinity the number $N$ of these ordered angles $\theta_k$, so that they cover the whole circle.  The discontinuities will go to zero in that limit, but this limit is not smooth, because of the rational numbers involved (see next sections).
\hypertarget{sub-the-periodicity-equation}{}
\subsection{The periodicity equation}
The periodicity condition \eqref{periodicity} for $n_\mu(\theta)$ demands that this random walk comes back after some number of steps $N\to \infty$. This is a nonlinear equation, involving $\beta $ plus $3 N$ random parameters $\vec r_l$ from all $N$ steps.
\begin{smalleq}
   && \Pi(\beta, r_.) = \prod_{ l =1}^N \left(\cos\beta I + \I \sin\beta \hat \tau \cdot \vec r_l \right) = I
\end{smalleq}
Each matrix factor in this product can be written as
\begin{smalleq}
   && \left(\cos\beta I + \I \sin\beta \hat \tau \cdot \vec r_l \right) = \hat P^+_l e^{\I \beta} + \hat P^-_l e^{-\I \beta};\\
   && \hat P^\pm_l = \frac{I \pm \hat \tau \cdot \vec r_l}{2};
\end{smalleq}
which results in the product being the polynomial of $e^{\pm\I \beta}$
\begin{smalleq}
   && \Pi_N(\beta, r_.) =\sum\displaylimits_{\sigma_1\dots\sigma_N= \pm 1} \exp{\displaystyle\I \beta \sum_{l =1}^N \sigma_l} \prod_{l=1}^N \hat P^{\sigma_l}_l
\end{smalleq}
This equation relates the last $SU_2 $ matrix factor to the product of the previous factors
\begin{smalleq}
    \cos\beta I + \I \sin\beta \hat \tau \cdot \vec r_N = \Pi^\dag_{N-1}(\beta,r_.)
\end{smalleq}
This matrix equation leads to a transcendental equation for $\beta$ given unit vectors $\vec r_1, \dots, \vec r_{N-1}$
\begin{smalleq}\hypertarget{eqbeta}{}
\label{eqbeta}
    \cos\beta = \oh \tr \Pi^\dag_{N-1}(\beta,r_.)
\end{smalleq}
Once $\beta$ satisfies this equation, the last vector $\vec r_N$ is fixed as a function of previous vectors and $\beta$
\begin{smalleq}
    \vec r_N = \frac{\tr \hat \tau \Pi^\dag_{N-1}(\beta,r_.)}{2 \sin\beta}
\end{smalleq}
The matrices in $\Pi$ all belonging to $SU_2 $, it is guaranteed that the right side of \eqref{eqbeta} is a real number between $ -1$ and $ 1$.

This makes it a plausible conjecture that at least for some values of $\beta$ there is a set of vectors $\vec r_1,\dots , \vec r_N$ satisfying this equation.

In the next sections, we consider a specific solution to this equation and its deformation to a general solution.

\hypertarget{sub-random-walk-on-regular-star-polygons}{}
\subsection{Random walk on regular star polygons}
\pdfbookmark[2]{Random walk on regular star polygons}{Euler}

Leaving a general study of this periodicity condition for a future mathematical work, we point out the special case of the random walk around a big circle on a sphere $\mathbb S_3$, studied in \cite{migdal2023exact, migdal2024quantum, migdal2025duality}.

In terms of the general solution described in the previous section, this corresponds to all vectors $\vec r_l$ pointing in the same direction, which can be chosen as 
\begin{smalleq}
    &&\vec r_l = \left\{0,0,\sigma_l\right\};\\
    &&\sigma_l = \pm 1;\\
   && \left(\cos\beta I + \I \sin\beta \hat \tau \cdot \vec r_k \right) =
   \begin{Vmatrix}
       e^{\I\beta\sigma_k} &0\\
       0& e^{-\I\beta\sigma_k}\\
   \end{Vmatrix}
\end{smalleq}

This particular solution is singled out by its extra invariance: with global parameters $\vec r_k$, the transformation from $\theta-$ to $\theta+$ is independent of $\theta$. All the rotation matrices are diagonal, and the angles $\sigma_k \beta$ add up.
It becomes a random walk on a circle, with equal steps forward or backward.

In that case, the solution for $f_\mu$ reads
\begin{smalleq}\hypertarget{Euler}{}
\label{Euler}
  && f(\theta) = R \hat \Omega \cdot \left \{ \cos \gamma(\theta),\sin\gamma(\theta),0,0\right\};\\
  && R = \frac{1}{2 \sin\beta/2};\\
  && \gamma( \theta) = \beta \sum_{\theta'< \theta}\sigma(\theta'); \textit{ with } \sigma(\theta) \in \mathbb Z_2;\\
  && \hat \Omega \in O(4)
\end{smalleq}
\hypertarget{sub-the-parameter-tuning-for-the-local-limit}{}
\subsection{The parameter tuning for the local limit}
\pdfbookmark[2]{The parameter tuning for the local limit}{PhiDef}

The periodicity of this solution requires two extra conditions
\begin{smalleq}
    && \beta = \frac{2 \pi p}{q};\textit{ with } p/q \in \mathbb Q;\\
    && \sum_\theta \sigma(\theta) = r q ; \textit{ with } r \in \mathbb Z
\end{smalleq}
The values $\theta \in (0, 2 \pi)$ are arbitrary, as long as they are ordered. In the limit $N \to \infty$ these values will cover uniformly the unit circle.

We cannot take this limit in the solution for the momentum loop, as there is no ordinary function with a discontinuity at every value of its argument. Such functions can exist only as distributions (see example in the papers \cite{migdal2024quantum, migdal2025duality}). 

The observable correlation functions in physical space, related to variations $\ff{\sigma_{\mu\nu}}$ of the loop functional, have the local limit in the sense of ordinary functions, not the distributions. 

The flow time scale $\alpha$  must be tuned for that local limit, namely $\alpha \propto 1/N^2 \to 0$. This coefficient $\alpha$ plays the same role as viscosity $\nu$ in the \NS{} equation, and the limit $\alpha\to 0$ is analogous to the turbulent limit $\nu \to 0$.
We shall discuss this local limit below in more detail.

We are treating this curve as a limit of a random walk on a regular star polygon, with vertices (in complex notation for planar vectors)
\begin{smalleq}
    &&V_n = R(\beta)e^{\I \beta n};\\
    && f(\theta_k) = \hat \Omega\cdot \left\{\Re V_{n(k)} , \Im V_{n(k)}, 0, 0 \right\};\\
    && \theta_k = \frac{2 \pi k}{N};\\
    && n(k) = \sum_{l< k}\sigma_l;\textit{ with }\sigma_k = \pm 1;
\end{smalleq}
\hypertarget{sub-the-euler-ensemble}{}
\subsection{The Euler ensemble}
This set of rational and integer numbers $\frac{p}{q} \in \mathbb Q, \sigma_1\dots \sigma_N \in \mathbb Z_2, r \in \mathbb Z $ constrained by $\sum \sigma_k = r q$ was called the Euler ensemble in the first paper \cite{migdal2023exact}.
The Euler ensemble has a remarkable statistical distribution in the limit $N \to \infty$, related to the totient summatory function
\begin{smalleq}
    \hypertarget{PhiDef}{}
\label{PhiDef}
    \Phi(q) = \sum_{n=1}^q \varphi(n)
\end{smalleq}
The distribution of the variable $X(p,q) =\cot(\pi p/q)^2/N^2 = \left( 4 R^2 -1\right)/N^2$ for large co-prime $1 \le p<q < N$ was studied in \cite{migdal2024quantum}, and it is a discontinuous piecewise power like distribution
 \begin{smalleq}\hypertarget{CotDist}{}
\label{CotDist}
   && f_X(X)= \left(1-\frac{\pi ^2}{675 \zeta (5)}\right)\delta(X) +\frac{\pi^3}{3} X\sqrt{X}\Phi\left(\floor*{\frac{1}{\pi \sqrt{X}}}\right);
\end{smalleq}
depicted in Figure.\ref{fig::PiPhi}.
\pctWPDF{0.5}{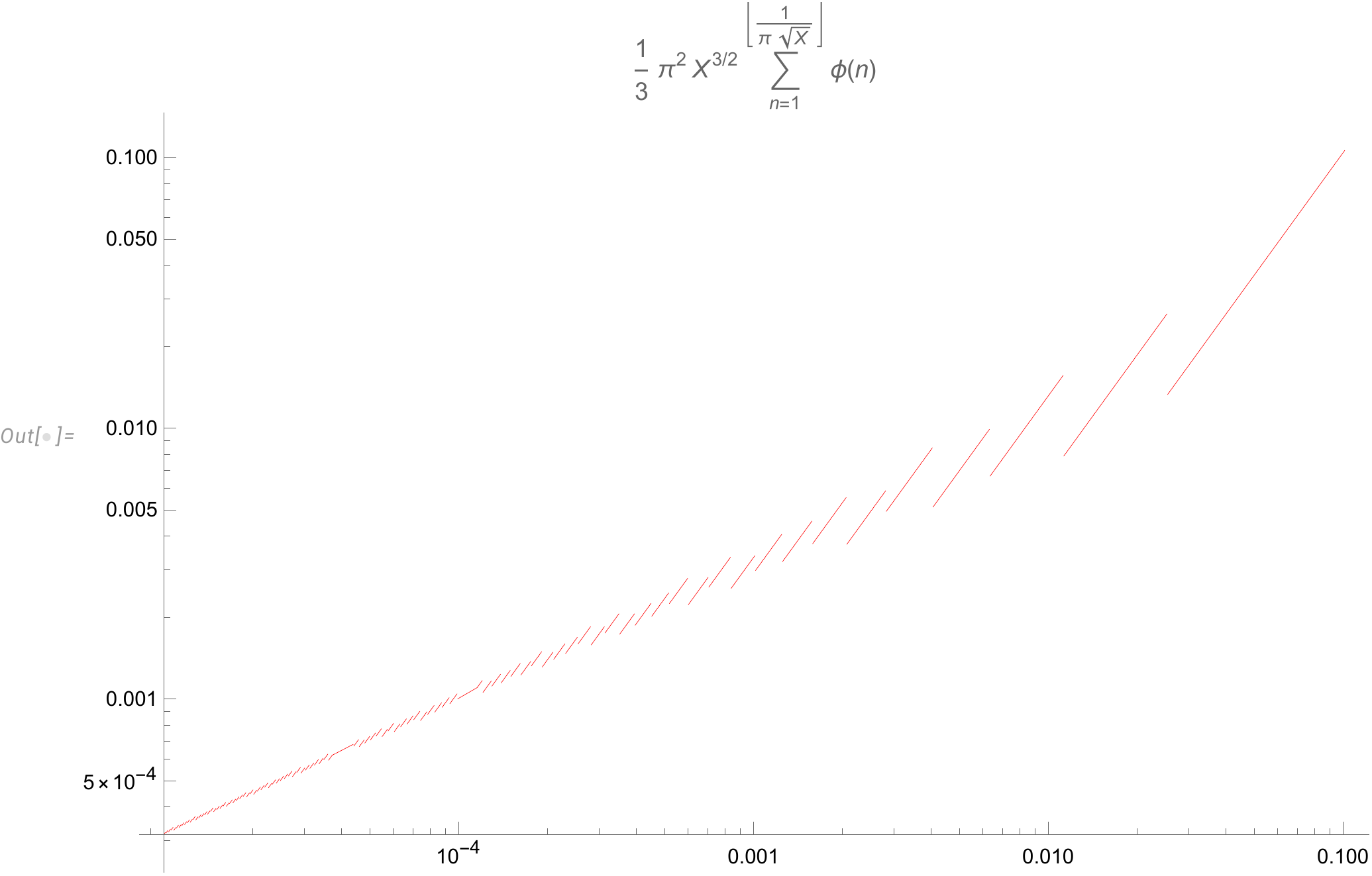}{Log log plot of the distribution \eqref{CotDist}}

\hypertarget{sub-the-continuum-limit}{}
\subsection{The continuum limit}
\pdfbookmark[2]{The continuum limit}{pathIntegralSol}

Substituting this solution into $W[C,\tau]$, and taking a local limit $N\to \infty$ we arrive at the following path integral in the  continuum limit
 \begin{subequations}\hypertarget{pathIntegralSol}{}
\label{pathIntegralSol}
          \begin{smalleq}
\hypertarget{WSol}{}
\label{WSol}
    &&W[C,\tau] = \nonumber\\
   &&\VEV{\exp{ \I \frac{ N \sqrt{X}}{2 \sqrt{\alpha (\tau+\tau_0)}}\int_0^{2\pi} d \theta \Im\left(  \dot{\mathcal{C}}_\Omega(\theta)
    e^{\imath\gamma(\theta)}\right)}}_{X,\gamma,\hat \Omega};\\
    && \mathcal C_\Omega(\theta) = C(\theta) \cdot \hat{\Omega} \cdot\{\imath,1,0,0\};\\
    \hypertarget{Xaverage}{}
\label{Xaverage}
    && \VEV{A[X,\gamma,\hat \Omega]}_{X,\gamma,\hat \Omega} =\nonumber\\
   &&
    \frac{\displaystyle\int_0^{\frac{1}{\pi^2}} d X\,f_X(X) \int\displaylimits_{\Omega \in O(4)} d \Omega\int [D \gamma] A[X,\gamma,\hat \Omega]}{\displaystyle|O(4)|\int [D\gamma]};\\
    &&  [D \gamma] = \int D \gamma(\theta) \exp{-\frac{N \pi X}{4}\int_0^{2 \pi} d \theta \gamma'^2};
\end{smalleq}
 \end{subequations}

We get the $U(1)$  statistical model with the boundary condition $\gamma(2\pi) = \gamma(0) + \beta N s$. The period $ \beta N s = 2 \pi p r$ is a multiple of $2 \pi$, which is irrelevant at $N \to \infty$. The effective potential for this theory is a linear function of the loop slope $\vec{C}'(\xi)$.
\hypertarget{sub-instanton-in-the-path-integral}{}
\subsection{Instanton in the path integral}
\pdfbookmark[2]{Instanton in the path integral}{classEq}

This classical equation for our path integral reads (with $\Omega \in O(4)$ being a random rotation matrix):
\begin{smalleq}
\hypertarget{classEq}{}
\label{classEq}
   && \gamma'' =  - \imath \kappa \left(\dot{\mathcal{C}}_\Omega \exp{\imath \gamma} + (\dot{\mathcal{C}}_\Omega )^\star \exp{-\imath \gamma}\right) ;\\
   && \kappa = \frac{1}{2 \pi \sqrt{X} \sqrt{\alpha (\tau + \tau_0) }};\\
   && \mathcal C_\Omega(\theta) =  \vec{C}(\theta) \cdot \hat{\Omega} \cdot\{\imath,1,0,0\};
\end{smalleq}
The parameter $\kappa$ is distributed according to the distribution $f_X(X)$ of the variable $X$ in a small Euler ensemble in the statistical limit.

This complex equation leads to a complex classical solution (instanton).
It simplifies for $ z = \exp{\imath \gamma}$:
\begin{smalleq}
    &&z'' = \frac{(z')^2}{z} + \kappa \left(\dot{\mathcal{C}}_\Omega z^2 + (\dot{\mathcal{C}}_\Omega )^\star \right);\\
    && z(0) = z(1) =1
\end{smalleq}
This equation cannot be analytically solved for arbitrary periodic function $\dot C_\Omega(\xi)$.

The weak and strong coupling expansions by $\kappa$ are straightforward.
At small $\kappa$
    \begin{smalleq}
    &&z(\theta) \to 1 + 2 \kappa \left(- A \theta +\int_0^\theta \Re \mathcal C_\Omega(\theta') d \theta' \right) + O(\kappa^2);\\
    && A = \int_0^{2\pi} \Re \mathcal C_\Omega(\theta')\frac{d \theta'}{2 \pi}
\end{smalleq}
At large $\kappa$
\begin{smalleq}
    &&z(\theta) \to \imath \exp{ -\imath \arg \dot{\mathcal{C}}_\Omega(\theta)} = \imath \frac{\abs{\dot{\mathcal{C}}_\Omega(\theta)}}{\dot{\mathcal{C}}_\Omega(\theta)}
\end{smalleq}
This solution is valid at intermediate $\theta$, not too close to the boundaries $\theta = (0,2\pi)$.
In the region near the boundaries $\theta(2\pi-\theta) \ll \frac{1}{\sqrt{\kappa}}$, the following asymptotic agrees with the classical  equation 
\begin{smalleq}
\hypertarget{endptZ}{}
\label{endptZ}
    &&z \to  1 \pm \imath\theta \frac{\sqrt{2 \kappa\Re \dot C_\Omega(0)}}{2 \pi} + O(\theta^2 );\\
     &&z \to  1 \pm \imath (2\pi-\theta) \frac{\sqrt{2 \kappa\Re \dot C_\Omega(2\pi)}}{2 \pi} + O((2\pi-\theta)^2);
\end{smalleq}
One can expand in small or large values of $\kappa$ and use the above distributions for $X$ term by term.

The classical limit of the circulation in exponential of \eqref{pathIntegralSol}
\begin{smalleq}
 \int_0^{2\pi} d \theta\Im\left( \dot{\mathcal{C}}_\Omega(\theta) \exp{\imath\gamma(\theta)} \right)\to \int_0^{2\pi} d \theta \abs{\dot{\mathcal{C}}_\Omega(\theta) }
\end{smalleq}
becomes a positive definite function of the rotation matrix $\Omega$. At large $\kappa$ the leading contribution will come from the rotation matrix minimizing this functonal.

Let us think about the physical meaning of this finding. We have just found the density of our Ising variables on a parametric circle
\begin{smalleq}
     \gamma(\theta) = \frac{\pi}{2} - \arg \dot{\mathcal{C}}_\Omega(\theta)
\end{smalleq}
This density does not fluctuate in a turbulent limit, except near the endpoints $\theta \to 0, \theta \to 2\pi$. In the vicinity of the endpoints, there is a different asymptotic solution \eqref{endptZ} for $ \gamma \to (z-1)/\imath$.

Computing the Wilson loop for a specific loop, say, the circle, is an interesting problem, but there is a simpler quantity.
In the papers \cite{migdal2024quantum,migdal2025mixing} we are considering an important calculable case of the vorticity correlation function, where the full solution in quadratures is available. 
This function has been directly observed in grid turbulence experiments \cite{GridTurbulence_1966} more than half a century ago and is being studied in modern large-scale real and numerical experiments\cite{SreeniDecaying, GDSM24, GregXi2}. 

The Euler ensemble solution \cite{migdal2024quantum} perfectly matches these real and numerical experiments in hydrodynamic turbulence, unlike old Kolmogorov laws, which strongly contradict this data.

These computations apply to the present four-dimensional \YM{} theory without much change. However, the random walk on regular star polygons is not the most general solution of MLE for the discrete manifold. As we have seen, the general solution corresponds to a random walk on the group $SU_2 $ with two random angular parameters at each step.

In the next section, we consider deforming this regular star polygon to achieve this general random walk with the built-in periodicity.

\hypertarget{sub-deformation-of-regular-polygons}{}
\subsection{Deformation of regular polygons}
\pdfbookmark[2]{Deformation of regular polygons}{Minimal}

Let us come back to the nonplanar random walk on a sphere, representing the general solution to the MLE in our theory. Such a solution was not present in the \NS{} case, because in that case, there was an extra term in the MLE, and this term was complex. The solution for $ F (\theta)$ had to be real, up to the constant vector term, independent of $\theta$. Such a constant vector term in $\vec F (\theta)$ does not contribute to the closed loop integral $\oint d\theta \dot C_\mu(\theta) F_\mu (\theta) $, keeping this integral real.
Starting with the constant imaginary part of $\vec F(\theta)$, the equations were solved unambiguously in \cite{migdal2025duality}.

In our case, there are fewer equations. Therefore, the most general solution is a random walk on a sphere, with two angular parameters at every step and a periodicity condition, which relates these parameters to $\beta$. There was no apparent way to directly solve this periodicity condition.

However, there is an indirect way to redefine the whole ensemble so that the periodicity is built into the solution.
Namely, consider the Euler ensemble of the previous section, with quantized $\beta = 2 \pi p/q$ and $\sigma_i = \pm 1$, and deform the corresponding closed path on the sphere $\mathbb S_3$ by shifting each vector $\vec n_k$  by some vector $\xi_k \in \mathbb R_4$. These shifts must be constrained so that the shifted vector stays on the unit sphere, plus the dot product between neighbors is $\cos \beta$, as in the Euler ensemble.
`
\begin{smalleq}
    &&\vec f_k = R \hat \Omega \cdot (\vec n^{0}_k + \vec \xi_k);\\
    && \hat \Omega \in SO(4);\\
    && \vec \xi_{N+1} = \vec \xi_1;\\
    && \vec n^0_k =\left \{ \cos \gamma_k, \sin\gamma_k, 0, 0\right\};\\
    && \gamma_k = \beta \sum_{l < k}\sigma_l;\\
    && \vec \xi_k^2 + 2 \vec \xi_k\cdot \vec n^{0}_k  =0;\\
    && \vec \xi_k\cdot \vec \xi_{k+1} + \vec \xi_k\cdot \vec n^{0}_{k+1} + \vec \xi_{k+1}\cdot \vec n^{0}_{k}=0;
\end{smalleq}
Our manifold (deformed Euler ensemble) is a set of $2N$ independent variables in $\vec \xi_k \in \mathbb  R_4$ left after the constraints. This is the same number of parameters per step as in our random walk on $SU_2 $ matrices.

There are no more periodicity conditions; they are trivially satisfied by the periodicity of the index $\vec \xi_{N+1} = \vec \xi_1$.
The invariant measure on this manifold  can be computed starting from the original invariant measure
\begin{smalleq}
   \int d \mu(n_.)= \prod_k \int d^4 n_k \delta(n_k^2-1) \delta(n_k\cdot n_{k+1} - \cos\beta)
\end{smalleq}
and changing variables $ n_k = n^0_k+ \xi_k$. This yields the following measure for $\xi_k$
\begin{smalleq}
   && \int d \mu(\xi_.) = \prod_k\int  d^4 \xi_k \delta\left(\vec \xi_k^2 + 2 \vec \xi_k\cdot \vec n^{0}_k\right) \nonumber\\
   && \delta\left(\vec \xi_k\cdot \vec \xi_{k+1} + \vec \xi_k\cdot \vec n^{0}_{k+1} + \vec \xi_{k+1}\cdot \vec n^{0}_{k}\right)= \nonumber\\
   && \prod_k \int d^4 \xi_k d u_k d v_k \exp{\I u_k \left(\vec \xi_k^2 + 2 \vec \xi_k\cdot \vec n^{0}_k\right) }\nonumber\\
   && \exp{\I v_k \left(\vec \xi_k\cdot \vec \xi_{k+1} + \vec \xi_k\cdot \vec n^{0}_{k+1} + \vec \xi_{k+1}\cdot \vec n^{0}_{k}\right)}
\end{smalleq}

Let us now put together the solution for the Wilson loop for the decaying solution of the \YM{} gradient flow
\begin{smalleq}
   && W(C,\tau) = \nonumber\\
   && \VEV{U(C,\tau)\exp{ \I M(X,\tau) \oint d \theta \Im\left( \dot{\mathcal{C}}_\Omega(\theta)
    e^{\imath\gamma(\theta)}\right)}}_{X,\gamma,\hat \Omega};\\
    &&U(C,\tau) = \nonumber\\
   && \int d\mu(\xi_.) \exp{\displaystyle\I M(X,\tau)\oint d \theta \dot C(\theta)\cdot \hat \Omega \cdot  \xi(\theta)};\\
   &&  \xi(\theta) = \xi_k \; \forall {\theta_k < \theta < \theta_{k+1}}
\end{smalleq}
Averaging over  $\VEV{}_{X,\gamma,\hat \Omega}$ is defined above, in \eqref{Xaverage}.

We keep the $N$ finite, but the Euler ensemble is now used in the statistical limit. Let us see what happens in the statistical limit with the ensemble of the deviations $\xi$ from the planar random walk on regular star polygons.

By rescaling $\vec\xi_k = \vec \eta_k/N$, we observe that at $ N\to\infty$ we can linearize the space of these shifts. The general solution of linearized constraints is an arbitrary vector orthogonal to the $x y$ plane
\begin{smalleq}
    \vec \eta_k = \left\{0,0,\vec \rho_k\right\} + O(1/N^2)
\end{smalleq}
This, however, leads to the singular solution, with the constraint that the whole loop $C$ lies in a plane
\begin{smalleq}
  &&  U(C,\tau) \propto \delta\left[\left(\vec C(.) \cdot \hat \Omega\right)_\perp\right];
\end{smalleq}

This solution is unphysical!

The only alternative solution is the pure random walk on regular star polygons, without any shifts; this corresponds to a special case
\begin{smalleq}
    \int d\mu(\xi_.) = \int \prod_k  d^3 \xi_k \delta(\vec \xi_k)
\end{smalleq}

The emergence of the universal statistical distribution, with a random walk on regular star polygons, is an unexpected link between mathematical physics, number theory, statistical mechanics, and group theory.

\hypertarget{sub-summary-of-the-decaying-gradient-flow}{}
\subsection*{Summary of the decaying gradient flow}
We can summarize what we have found for decaying \YM{} gradient flow as follows.
\begin{itemize}
    \item We found an exact statistical solution for the Wilson loop in the \YM{} gradient flow, approaching the pure gauge field fixed point as the inverse time $F_{\mu\nu} \propto 1/\tau$.
    \item This solution has the form of a universal statistical ensemble, parametrized by the rational numbers $\frac{r}{q} \in \mathbb Q$ and $N$ Ising spin variables $\sigma_k = \pm 1, k = 1,\dots, N$.
    \item These parameters (Euler ensemble) describe a periodic random walk on regular star polygons.
    \item On top of these, number theory parameters, there could also be $4N$ continuous parameters $ \xi_k \in \mathbb R_4, k = 1,\dots, N$,  subject to two constraints each.
    \item These $ 2 N$ free parameters describe periodic random walk on $SU_2 $, coupled to the random walk on regular star polygons.
    \item This extension of the Euler ensemble, however, appeared to have a singular continuum limit $N \to \infty$, enforcing a planar loop $C$, whose restriction is not acceptable as a solution for the Wilson loop of a gradient flow.
    \item This is a nontrivial statement for a compact manifold such as $SU_2 $, which is expected to be uniformly covered by a long enough random walk.
    \item This brings us back to the Euler ensemble as the only acceptable decaying solution of the \YM{} gradient flow.
    \item This solution manifests spontaneous quantization of a classical gradient flow, similar to the spontaneous emergence of the decaying turbulence in the deterministic \NS{} equation without any random forces.
\end{itemize}
In the forthcoming publication, we consider an even more interesting case of spontaneous quantization-- the nontrivial fixed point of the gradient flow.

\hypertarget{sec-discussion-emerging-quantum-world}{}
\section{Discussion: Emerging Quantum World}

We have identified two \emph{exact} quantum solutions emerging from a \emph{classical}
gradient flow in arbitrary non-Abelian gauge theories. In principle, these solutions
can describe confined gauge fields and quarks in QCD.

This is \emph{not} conventional QCD: we study the pure gradient flow, without adding
stochastic forces. Adding white noise to the gradient of the Yang--Mills (YM) action
yields the Langevin equation for the \emph{quantized} field theory; the fixed point
of that stochastic process reproduces the functional integral with Gibbs weight
$\exp{-S_{\mathrm{YM}}}$.

\subsection*{Langevin equation and asymptotic freedom}

Let us briefly recall stochastic quantization. (We stress that we are \emph{not}
solving full QCD here; that will be the subject of a separate paper once remaining
technical points are settled.)

Formally, asymptotically free QCD corresponds to the $g_0\to 0$ limit of the fixed
point of the Langevin equation:
\begin{smalleq}
    &&\pd{\tau} A_\nu =  D_\mu F_{\mu\nu} + f_\nu;\\
    && \VEV{f_\mu(x,\tau) \otimes f_\nu(y,\tau')} = g_0^2 \delta(\tau-\tau')\delta_{\mu\nu}\delta^4(x-y) \hat \Pi;\\
    && \Pi^{k l}_{i j} = \delta_{i j} \delta_{k l} - \frac{1}{N_c} \delta_{i l} \delta_{k j}
\end{smalleq}
Stochastic quantization avoids several ambiguities of the formal 4D path integral and
is a convenient route to the planar ($N\to\infty$) limit.

The Wilson loop then satisfies the loop equation with a contact term
\cite{MMEq79,Mig83}:
\begin{smalleq}\label{loopEq}
    && \pd{\tau}  W[C] = \int_0^{2\pi} d \theta  \dot C_\nu(\theta) \partial_\mu(\theta)\fbyf{W[C]}{\sigma_{\mu\nu}(\theta)}\nonumber\\
    &&+ g_0^2 \int_0^{2\pi} d \theta_2  \dot C_\mu(\theta_2) \int_0^{\theta_2} d \theta_1 \dot C_\mu(\theta_1) \delta^4\left(C(\theta_1) - C(\theta_2)\right)\nonumber\\
   &&\left(W\left[C_{\theta_1,\theta_2} \uplus C_{\theta_2,\theta_1}\right] - \frac{W[C]}{N} \right);\\
   && 
   \begin{cases}
    C_{\theta_1,\theta_2}\left(\frac{2\pi(\theta - \theta_1)}{\theta_2-\theta_1}\right) = C(\theta)& \textit{ if } \theta_1 < \theta < \theta_2\\
        C_{\theta_2,\theta_1}\left(\frac{2\pi(\theta - \theta_2)}{2 \pi+\theta_1-\theta_2} \right)=  C(\theta)& \textit{ if } \theta_2 < \theta < \theta_1 + 2 \pi\\
   \end{cases}
\end{smalleq}
Geometrically, $C_{\theta_1,\theta_2} \uplus C_{\theta_2,\theta_1}$ is a pair of
closed loops obtained by cutting the original loop at $\theta_{1,2}$ and closing the
arcs; in our functional analysis this is a special case of a periodic map
$C_\alpha(\theta)$ with a smaller intrinsic period inside $[0,2\pi]$.

In the large-$N$ (planar) limit the contact term factorizes \cite{MMEq79}:
\begin{smalleq}\label{MMEq}
    && \pd{\tau}  W[C] =  \int_0^{2\pi} d \theta  \dot C_\nu(\theta) \partial_\mu(\theta)\fbyf{W[C]}{\sigma_{\mu\nu}(\theta)}\nonumber\\
    &&+ \lambda \int_0^{2\pi} d \theta_2  \dot C_\mu(\theta_2) \int_0^{\theta_2} d \theta_1 \dot C_\mu(\theta_1) \delta^4\left(C(\theta_1) - C(\theta_2)\right)\nonumber\\
    &&W\left[C_{\theta_1,\theta_2}\right] W\left[C_{\theta_2,\theta_1}\right];\\&& \lambda = N g_0^2 \to \const{}
\end{smalleq}

\subsection*{QCD as a fixed point of \YM{} gradient flow?}

The QCD Wilson loop corresponds to the fixed point of \eqref{MMEq}, with the limit
$\lambda\to 0$ taken together with $\Lambda\to\infty$ at fixed
$\mu=\Lambda\,\lambda^b \exp{-\frac{a}{\lambda}}$ (with the standard RG coefficients
$a,b$). The cutoff $\Lambda$ enters through a gauge-invariant regularization of the
contact term (e.g.\ replacing $\delta^4(x-y)$ by a narrow Gaussian of variance
$\sigma=1/\Lambda$ and inserting ordered path exponentials to preserve gauge
invariance). These details are inessential for us: in momentum loop space they drop
out \cite{MLDMig86,Mig98Hidden}.

The noise amplitude in the Langevin equation is proportional to the bare coupling,
which vanishes as the inverse logarithm of the UV cutoff (or lattice spacing).
Formally, in the local limit the stochastic force vanishes and QCD reduces to the
\emph{fixed point} of the gradient flow. The dynamical question is whether the limits
$\tau\to\infty$ (stochastic time) and $g_0\to 0$ commute; on the lattice, one takes
$\tau\to\infty$ first, then $g_0\to 0$.

Here we explore a different route: we keep the \emph{continuum} loop equation for the
gradient flow, without discretizing spacetime. This preserves translation and
rotation invariance in 4D and reveals a rich family of \textbf{analytic solutions}
in momentum loop space.

\subsection*{A rich family of fixed points of 4D \YM{} flow (does it include QCD?)}

Given the correspondence between gradient-flow fixed points and the weak-coupling
limit of the Langevin fixed point, we argue that \textbf{if there is a family of
gradient-flow fixed points} for Wilson loops, then asymptotically free QCD should
select a \emph{subset} of that manifold in a well-defined local limit.

The key question is whether any of the self-dual fixed points identified above match
asymptotically free QCD in such a limit, and whether that limit exists at all. We
will address this in a forthcoming paper. If the family and subset do exist, then
quantization of YM could \emph{emerge spontaneously}, much as turbulent statistics
emerge from deterministic Navier--Stokes dynamics. 
\subsection*{Topological vacua and a singular zero-temperature limit of 4D statistics}
Among these fixed points, the
self-dual ones are the only solutions for which stability is rigorously controlled,
via the standard inequality bounding the YM action (interpreted as 4D energy in the
Gibbs weight) by the absolute value of the topological charge.

The spectrum of small fluctuations around a self-dual minimum is positive, implying
exponential decay with stochastic time $\tau$; thus self-dual solutions of
\eqref{LoopEq} are attractors of the YM gradient flow. As discussed above, the
low-temperature limit of 4D YM selects energy minima that remain at \textbf{finite}
distance from the vacuum (do not scale with volume $V\to\infty$). Any configuration
with finite topological charge $Q$ satisfies this, since its energy is $|Q|$.

Therefore the low-temperature partition function must be a superposition over
degenerate topological vacua. Symbolically,
\begin{smalleq}
W(C,\infty)\big|_{\text{QCD}} \sim \sum_{Q = -\infty}^{\infty} \exp{-\frac{|Q|}{g_R^2}} W(C,\infty)\Big|_{\int F \wedge \star F = Q},
\end{smalleq}
where $g_R^2$ is the effective coupling renormalized by fluctuations around an
instanton at the QCD scale. A classic example of such instanton sums is the work of
Nekrasov \cite{nekrasov2003SYMinstantons} on the Seiberg--Witten prepotential for
$\mathcal{N}=2$ SYM.
At nonzero temperature, periodic instanton (‘caloron’) sectors `a la van Baal \cite{Caloron2004} provide the finite-T analogues of the self-dual fixed points considered here; analyzing their role in the loop-space framework is an interesting direction for future work.
\subsection*{Old conjectures about spontaneous quantization of gauge theory}

This picture echoes the conjecture in \cite{Migdal_1986_Stochastic} that quantum
mechanics and QFT can emerge from chaotic nonlinear dynamics in an auxiliary time,
without explicit noise: the Langevin equation quantizes a gauge theory, and its
zero-noise limit reduces to a gradient flow. That speculative idea finds concrete
support here in the long-time behavior of the YM gradient flow, linking
deterministic classical dynamics to quantum behavior and suggesting a path toward the
foundations of quantum theory and nonperturbative gauge dynamics.\footnote{The
concluding section of \cite{Migdal_1986_Stochastic} states: ``The Langevin equation
offers us a new possibility of generalizing quantum field theory so that it can be
`derived' in some approximation from more fundamental laws. Of course, here we enter
the world of fantasy, but after the advances made in quantum field theory during the
last few years, this fantasy may serve as a stimulus to the creation of realistic
models. The basic immutable law of quantum theory is the principle of superposition,
according to which each process is characterized by a complex amplitude that is the
sum of the amplitudes of the alternative histories of the process. On the other
hand, in Nature, linear processes are the idealizations, and actual phenomena are
often nonlinear. Is this the case in quantum theory? What if quantum theory is a
linearization of the equations of some more fundamental theory, acting over
distances of the order of the Planck length? We shall never know unless we try to
construct at least a rough model of this type of theory. The Langevin equation
presents us with a natural basis for this. In particular, we may suppose that this
equation is an approximation to a nonlinear dynamic system in which $T$ (or $f$ in
Minkowski space) plays the part of time.''}

\paragraph{Concluding remark (Self-organized quantization).}
Our construction provides, to our knowledge, a semiclassical (WKB) string functional
that solves the gauge-theory loop equation directly in four spacetime dimensions,
without extra compact directions. The gradient-flow time $\tau$ labels a family of
configurations and is not an observable at fixed $\tau$. By \emph{self-organized
quantization} we mean: starting from a classical field, the flow spreads over
configuration space, may branch, and evolves into a manifold of trajectories endowed
with a universal probability distribution; depending on initial data it either covers
a topological fixed manifold or approaches a pure-gauge vacuum along a universal
trajectory---described by another manifold (Euler ensemble). In this setting, linear
superposition is an \emph{exact} property of the gradient flow viewed as an evolution
of probability distributions: from a delta-like classical state to a broad quantum
distribution. Thus a quantum Wilson-loop functional can emerge from the turbulent
dynamics of the \YM{} flow.

Beyond the technical achievement of exact solutions, this perspective reveals an
unexpected link between nonlinear flows and loop-space diffusion, connecting PDEs to
geometry, statistics, and number theory in a spirit faintly reminiscent of the
Langlands program.

\smallskip
\noindent\textit{Outlook.}
A key next step is to find fixed-point solutions of the planar QCD loop equation
including the contact terms; we will report on this elsewhere.

\setcounter{equation}{0}

\hypertarget{sec-acknowledgements}{}
\section*{Acknowledgements}
I benefited from discussions with Camillo de Lellis, Semon Rezchikov, and Juan Maldacena.
I am especially grateful to Edward Witten for his critical insights, 
which helped me to sharpen the formulation of my results and to clarify their broader significance.
 
This research was supported by the Simons Foundation award ID SFI-MPS-T-MPS-00010544 in the Institute for Advanced Study.

\hypertarget{sec-declaration-of-generative-ai-and-ai-assisted-technologies-in-the-writing-process}{}
\section*{Declaration of generative AI and AI-assisted technologies in the writing process}
During the preparation of this work, the author used ChatGPT5 in order to fix typos and improve style. After using this tool, the author reviewed and edited the content as needed and takes full responsibility for the content of the publication.

\bibliographystyle{elsarticle-num}
\bibliography{bibliography}

\begin{thebibliography}{10}
\expandafter\ifx\csname url\endcsname\relax
  \def\url#1{\texttt{#1}}\fi
\expandafter\ifx\csname urlprefix\endcsname\relax\def\urlprefix{URL }\fi
\expandafter\ifx\csname href\endcsname\relax
  \def\href#1#2{#2} \def\path#1{#1}\fi

\bibitem{Feehan2016}
P.~M.~N. Feehan, \href{https://arxiv.org/abs/1409.1525}{Global existence and convergence of solutions to gradient systems and applications to yang-mills gradient flow} (2016).
\newblock \href {http://arxiv.org/abs/1409.1525} {\path{arXiv:1409.1525}}.
\newline\urlprefix\url{https://arxiv.org/abs/1409.1525}

\bibitem{Waldron2019}
A.~Waldron, \href{http://dx.doi.org/10.1007/s00222-019-00877-2}{Long-time existence for yang–mills flow}, Inventiones mathematicae 217~(3) (2019) 1069–1147.
\newblock \href {https://doi.org/10.1007/s00222-019-00877-2} {\path{doi:10.1007/s00222-019-00877-2}}.
\newline\urlprefix\url{http://dx.doi.org/10.1007/s00222-019-00877-2}

\bibitem{migdal2023exact}
A.~Migdal, \href{http://dx.doi.org/10.3390/fractalfract7100754}{To the theory of decaying turbulence}, Fractal and Fractional 7~(10) (2023) 754.
\newblock \href {http://arxiv.org/abs/2304.13719} {\path{arXiv:2304.13719}}, \href {https://doi.org/10.3390/fractalfract7100754} {\path{doi:10.3390/fractalfract7100754}}.
\newline\urlprefix\url{http://dx.doi.org/10.3390/fractalfract7100754}

\bibitem{migdal2024quantum}
A.~Migdal, Quantum solution of classical turbulence: Decaying energy spectrum, Physics of Fluids 36~(9) (2024) 095161.
\newblock \href {https://doi.org/10.1063/5.0228660} {\path{doi:10.1063/5.0228660}}.

\bibitem{MMEq79}
Y.~Makeenko, A.~Migdal, \href{url{https://www.sciencedirect.com/science/\\ article/pii/037026937990131X}}{Exact equation for the loop average in multicolor qcd}, Physics Letters B 88~(1) (1979) 135--137.
\newblock \href {https://doi.org/url{https://doi.org/10.1016/0370-2693(79)90131-X}} {\path{doi:url{https://doi.org/10.1016/0370-2693(79)90131-X}}}.
\newline\urlprefix\url{url{https://www.sciencedirect.com/science/\\ article/pii/037026937990131X}}

\bibitem{MM1981NPB}
Y.~M. Makeenko, A.~A. Migdal, Quantum chromodynamics as dynamics of loops, Nuclear Physics B 188 (1981) 269--316.
\newblock \href {https://doi.org/10.1016/0550-3213(81)90105-2} {\path{doi:10.1016/0550-3213(81)90105-2}}.

\bibitem{Mig83}
A.~Migdal, { Loop equations and $\inv{N}$ expansion}, Physics Reports 201 (1983).

\bibitem{M23PR}
A.~Migdal, \href{https://arxiv.org/abs/2209.12312}{Statistical equilibrium of circulating fluids}, Physics Reports 1011C (2023) 1--117.
\newblock \href {http://arxiv.org/abs/2209.12312} {\path{arXiv:2209.12312}}, \href {https://doi.org/10.48550/ARXIV.2209.12312} {\path{doi:10.48550/ARXIV.2209.12312}}.
\newline\urlprefix\url{https://arxiv.org/abs/2209.12312}

\bibitem{migdal2025duality}
A.~Migdal, \href{https://doi.org/10.1142/S0217751X25410040}{Duality of navier–stokes to a one-dimensional system}, International Journal of Modern Physics A 0~(0) (2025) 2541004.
\newblock \href {http://arxiv.org/abs/https://doi.org/10.1142/S0217751X25410040} {\path{arXiv:https://doi.org/10.1142/S0217751X25410040}}, \href {https://doi.org/10.1142/S0217751X25410040} {\path{doi:10.1142/S0217751X25410040}}.
\newline\urlprefix\url{https://doi.org/10.1142/S0217751X25410040}

\bibitem{Mig98Hidden}
A.~A. Migdal, {Hidden symmetries of large N QCD}, Prog. Theor. Phys. Suppl. 131 (1998) 269--307.
\newblock \href {http://arxiv.org/abs/hep-th/9610126} {\path{arXiv:hep-th/9610126}}, \href {https://doi.org/10.1143/PTPS.131.269} {\path{doi:10.1143/PTPS.131.269}}.

\bibitem{Montgomery2002}
R.~Montgomery, A Tour of Subriemannian Geometries, Their Geodesics and Applications, Vol.~91 of Mathematical Surveys and Monographs, American Mathematical Society, 2002.

\bibitem{Osserman2002}
R.~Osserman, A Survey of Minimal Surfaces, reprint of the 2nd edition, 1986 Edition, Dover Publications, 2002, originally published by Van Nostrand Reinhold, 1969.

\bibitem{Maldacena1998}
J.~M. Maldacena, \href{https://arxiv.org/abs/hep-th/9711200}{The large n limit of superconformal field theories and supergravity}, Advances in Theoretical and Mathematical Physics 2 (1998) 231--252.
\newblock \href {http://arxiv.org/abs/hep-th/9711200} {\path{arXiv:hep-th/9711200}}.
\newline\urlprefix\url{https://arxiv.org/abs/hep-th/9711200}

\bibitem{Drukker_Gross_Ooguri1999}
N.~Drukker, D.~J. Gross, H.~Ooguri, Wilson loops and minimal surfaces, Phys. Rev. D 60 (1999) 125006.
\newblock \href {http://arxiv.org/abs/hep-th/9904191} {\path{arXiv:hep-th/9904191}}, \href {https://doi.org/10.1103/PhysRevD.60.125006} {\path{doi:10.1103/PhysRevD.60.125006}}.

\bibitem{PolyakovRychkov}
A.~M. Polyakov, V.~S. Rychkov, Gauge fields -- strings duality and the loop equation, Nuclear Physics B 581 (2000) 116--134.
\newblock \href {http://arxiv.org/abs/hep-th/0002106} {\path{arXiv:hep-th/0002106}}, \href {https://doi.org/10.1016/S0550-3213(00)00177-9} {\path{doi:10.1016/S0550-3213(00)00177-9}}.

\bibitem{tHooft1976}
G.~'t~Hooft, Computation of the quantum effects due to a four-dimensional pseudoparticle, Physical Review D 14 (1976) 3432--3450, erratum: \emph{Phys. Rev. D} \textbf{18}, 2199 (1978).
\newblock \href {https://doi.org/10.1103/PhysRevD.14.3432} {\path{doi:10.1103/PhysRevD.14.3432}}.

\bibitem{Nambu1970}
Y.~Nambu, Duality and hydrodynamics, in: Lectures at the Copenhagen Symposium, 1970, reprinted in \emph{Broken Symmetry: Selected Papers of Y. Nambu}, eds. T. Eguchi and K. Nishijima, World Scientific, 1995, p. 280.

\bibitem{Goto1971}
T.~Goto, Relativistic quantum mechanics of one-dimensional mechanical continuum and subsidiary condition of dual resonance model, Progress of Theoretical Physics 46~(5) (1971) 1560--1569.
\newblock \href {https://doi.org/10.1143/PTP.46.1560} {\path{doi:10.1143/PTP.46.1560}}.

\bibitem{Morrey1948}
C.~B. Morrey, The problem of plateau on a riemannian manifold, Annals of Mathematics 49~(4) (1948) 807--851, discusses conformal parametrizations and boundary regularity for minimal surfaces in Riemannian manifolds.
\newblock \href {https://doi.org/10.2307/1969212} {\path{doi:10.2307/1969212}}.

\bibitem{ATMP1999}
J.~C. Baez, J.~W. Barrett, The quantum tetrahedron in 3 and 4 dimensions, Adv. Theor. Math. Phys. 3 (1999) 815--850.

\bibitem{Batista2013}
C.~Batista, Weyl tensor classification in four-dimensional manifolds of all signatures, arXiv preprint arXiv:1304.3771 (2013).

\bibitem{Douglas1931}
J.~Douglas, \href{https://doi.org/10.2307/1989631}{Solution of the problem of plateau}, Transactions of the American Mathematical Society 33~(1) (1931) 263--321.
\newblock \href {https://doi.org/10.2307/1989631} {\path{doi:10.2307/1989631}}.
\newline\urlprefix\url{https://doi.org/10.2307/1989631}

\bibitem{DualityEquations}
A.~Migdal, Duality equations for variousgroups, \url{https://www.wolframcloud.com/obj/sasha.migdal/Published/DualityEquationsForVariousGroups.nb} (07 2025).

\bibitem{M93}
A.~Migdal, \href{https://arxiv.org/abs/hep-th/9310088}{Loop equation and area law in turbulence}, in: L.~Baulieu, V.~Dotsenko, V.~Kazakov, P.~Windey (Eds.), Quantum Field Theory and String Theory, Springer {US}, 1995, pp. 193--231.
\newblock \href {https://doi.org/10.1007/978-1-4615-1819-8} {\path{doi:10.1007/978-1-4615-1819-8}}.
\newline\urlprefix\url{https://arxiv.org/abs/hep-th/9310088}

\bibitem{MLDMig86}
A.~Migdal, \href{https://www.sciencedirect.com/science/article/pii/0550321386903317}{Momentum loop dynamics and random surfaces in qcd}, Nuclear Physics B 265~(4) (1986) 594--614.
\newblock \href {https://doi.org/https://doi.org/10.1016/0550-3213(86)90331-7} {\path{doi:https://doi.org/10.1016/0550-3213(86)90331-7}}.
\newline\urlprefix\url{https://www.sciencedirect.com/science/article/pii/0550321386903317}

\bibitem{SecQuanM95}
A.~Migdal, \href{https://www.sciencedirect.com/science/article/pii/092056329500433A}{Second quantization of the wilson loop}, Nuclear Physics B - Proceedings Supplements 41~(1) (1995) 151--183.
\newblock \href {https://doi.org/https://doi.org/10.1016/0920-5632(95)00433-A} {\path{doi:https://doi.org/10.1016/0920-5632(95)00433-A}}.
\newline\urlprefix\url{https://www.sciencedirect.com/science/article/pii/092056329500433A}

\bibitem{nagasawa1989asymptotic}
T.~Nagasawa, On asymptotic stability of yang-mills' gradient flow, in: RIMS Kokyuroku, Vol. 698, 1989, pp. 171--187.

\bibitem{migdal2025mixing}
A.~Migdal, Dual theory of turbulent mixing, \url{https://arxiv.org/abs/2504.10205} (2025).
\newblock \href {http://arxiv.org/abs/2504.10205} {\path{arXiv:2504.10205}}.

\bibitem{GridTurbulence_1966}
G.~Comte-Bellot, S.~Corrsin, The use of a contraction to improve the isotropy of grid-generated turbulence, Journal of Fluid Mechanics 25~(4) (1966) 657–682.
\newblock \href {https://doi.org/10.1017/S0022112066000338} {\path{doi:10.1017/S0022112066000338}}.

\bibitem{SreeniDecaying}
J.~Panickacheril~John, D.~A. Donzis, K.~R. Sreenivasan, Laws of turbulence decay from direct numerical simulations, Philos. Trans. A Math. Phys. Eng. Sci. 380~(2218) (2022) 20210089.

\bibitem{GDSM24}
S.~Migdal, Decaying turbulence experiment, \url{https://drive.google.com/drive/folders/1DkHOOxhbsT0prVj65wJPnHOeb_EXc7N6?usp=drive_link}, accessed: 18-November-2024 (2024).

\bibitem{GregXi2}
C.~K\"uchler, G.~P. Bewley, E.~Bodenschatz, \href{https://link.aps.org/doi/10.1103/PhysRevLett.131.024001}{Universal velocity statistics in decaying turbulence}, Phys. Rev. Lett. 131 (2023) 024001.
\newblock \href {https://doi.org/10.1103/PhysRevLett.131.024001} {\path{doi:10.1103/PhysRevLett.131.024001}}.
\newline\urlprefix\url{https://link.aps.org/doi/10.1103/PhysRevLett.131.024001}

\bibitem{nekrasov2003SYMinstantons}
N.~A. Nekrasov, \href{https://arxiv.org/abs/hep-th/0306211}{Seiberg-witten prepotential from instanton counting} (2003).
\newblock \href {http://arxiv.org/abs/hep-th/0306211} {\path{arXiv:hep-th/0306211}}.
\newline\urlprefix\url{https://arxiv.org/abs/hep-th/0306211}

\bibitem{Caloron2004}
F.~Bruckmann, P.~van Baal, Calorons with non-trivial holonomy on and off the lattice, in: Proceedings of Lattice 2004, Vol. 140, 2005, pp. 494--501.
\newblock \href {http://arxiv.org/abs/hep-lat/0408030} {\path{arXiv:hep-lat/0408030}}.

\bibitem{Migdal_1986_Stochastic}
A.~A. Migdal, \href{https://dx.doi.org/10.1070/PU1986v029n05ABEH003373}{Stochastic quantization of field theory}, Soviet Physics Uspekhi 29~(5) (1986) 389.
\newblock \href {https://doi.org/10.1070/PU1986v029n05ABEH003373} {\path{doi:10.1070/PU1986v029n05ABEH003373}}.
\newline\urlprefix\url{https://dx.doi.org/10.1070/PU1986v029n05ABEH003373}

\bibitem{Duren2004}
P.~L. Duren, Harmonic Mappings in the Plane, Vol. 156 of Cambridge Tracts in Mathematics, Cambridge University Press, Cambridge, 2004.

\end{thebibliography}
\appendix
\hypertarget{sub-minimal-surfaces}{}
\section{Minimal surfaces}\hypertarget{Minimal}{}
\label{Minimal}

We briefly summarize, in a form convenient for field theorists, the
classical Weierstrass-Douglas theory of minimal surfaces\cite{Douglas1931,Osserman2002,Duren2004}. A surface is given
parametrically by
\begin{smalleq}
S: \ral = X_{\alpha}\left(\xi_1,\xi_2\right)
\end{smalleq}
and its area is obtained by minimizing
\begin{smalleq}
A[X] = \int_S \sqrt{d \sigma_{\mu\nu}^2} = \int d^2 \xi \sqrt{\mbox{Det } G}
\end{smalleq}
where the induced metric is
\begin{smalleq}
G_{ab} = \partial_a X_{\mu}\partial_b X_{\mu}.
\end{smalleq}

A convenient auxiliary formulation introduces a unit tangent 2–tensor as an
independent field and minimizes
\begin{smalleq}
A\left[X,t,\lambda\right] = \int d^2 \xi  \left(\,e_{ab} \partial_a X_{\mu}
\partial_b X_{\nu} \, t_{\mu\nu} + \lambda \left(1-  t_{\mu\nu}^2 \right) \right)
\end{smalleq}
The Euler–Lagrange equations imply
\begin{smalleq}
t_{\mu\nu} = \frac{e_{ab}}{2\lambda}  \partial_a X_{\mu} \partial_b X_{\nu}\;;
t_{\mu\nu}^2 = 1,
\end{smalleq}
which shows the equivalence to the original definition.

For practical computations, it is useful to introduce an internal metric
\(g_{ab}\) and consider the quadratic functional
\begin{smalleq}
A\left[X,g\right] = \oh \int_S d^2 \xi  \, \tr{ g^{-1} G} \,\sqrt{\mbox{Det }
g}.
\hypertarget{gG}{}
\label{gG}
\end{smalleq}
Stationarity with respect to \(g\) gives
\begin{smalleq}
 g_{ab} \,\tr{ g^{-1} G} =  2 G_{ab},
\end{smalleq}
whose solution set is
\begin{smalleq}
g_{ab} = \lambda G_{ab}.
\end{smalleq}
The local scale \(\lambda\) drops out of the area, so one may first minimize
\rf{gG} with respect to \(X\) (linear problem) and then eliminate \(g\)
(nonlinear step).

Choosing conformal coordinates diagonalizes the metric:
\begin{smalleq}
g_{ab} = \delta_{ab} \rho,\; g^{-1}_{ab} = \frac{\delta_{ab}}{\rho},\;
\sqrt{\mbox{Det } g} = \rho;
\end{smalleq}
after which the area becomes
\begin{smalleq}
A[X,\rho] = \oh \int_S d^2 \xi \partial_a X_{\mu} \partial_a X_{\mu}.
\end{smalleq}
The scale \(\rho\) remains implicit via boundary data.

To allow arbitrary boundary parametrizations, we take the upper half–plane
in \(\xi\), with boundary \(\xi_2=0\), and impose
\begin{smalleq}
X_{\mu}(\xi_1,+0) = C\left(f(\xi_1)\right),
\hypertarget{Bcon}{}
\label{Bcon}
\end{smalleq}
where the unknown boundary map \(f\) is tied to \(\rho\) by
\begin{smalleq}
g_{11}  = \rho =  G_{11} = \left(\partial_1 X_{\mu}\right)^2 = C_{\mu}'^2 f'^2
\end{smalleq}
Thus one solves the linear problem for \(X\) at fixed \(f\), evaluates the
area, and then minimizes over \(f(.)\). The corresponding extremality
condition coincides with the diagonality of the boundary metric:
\begin{smalleq}
\left[\partial_1 X_{\mu} \partial_2 X_{\mu}\right]_{\xi_2=+0} = 0
\hypertarget{Diag}{}
\label{Diag}
\end{smalleq}

The bulk equations are just Laplace’s equation \(\partial^2 X=0\) in the upper
half–plane with Dirichlet data \rf{Bcon}. The solution is the Poisson
integral
\begin{smalleq}
X_{\mu}(\xi) = \int_{-\8}^{+\8} \frac{d t}{\pi} \frac{C_{\mu}\left(f(t)\right)
\,
\xi_2}{\left(\xi_1-t\right)^2 + \xi_2^2}
\end{smalleq}
Using \(\partial^2 X=0\), the area reduces to a boundary term:
\begin{eqnarray}
&&A[f]  = \oh \int d^2 \xi \partial_a \left(X_{\mu} \partial_a X_{\mu}\right) =\nonumber\\
&& -\oh
\int_{-\8}^{+\8} d \xi_1 \left[X_{\mu} \partial_2  X_{\mu}\right]_{\xi_2 = +0}
\end{eqnarray}
Substituting the solution gives
\begin{eqnarray}
&&A[f] = -\frac{1}{2\pi}\, \Re \int_{-\8}^{+\8}d t \int_{-\8}^{+\8} d t'\nonumber\\
&&
\frac{C_{\mu}(f(t))\, C_{\mu}(f(t'))}{(t-t'-\i0)^2}
\end{eqnarray}
which can be recast as the manifestly positive form
\begin{smalleq}\hypertarget{Douglas}{}
\label{Douglas}
A[f] = \frac{1}{4\pi}\,  \int_{-\8}^{+\8}d t \int_{-\8}^{+\8} d t'
\frac{\left(C_{\mu}(f(t))- C_{\mu}(f(t'))\right)^2}{(t-t')^2}
\end{smalleq}
An integration by parts yields
\begin{eqnarray}
&&A[f] = \frac{1}{2\pi}\,  \int_{-\8}^{+\8}d t f'(t) \int_{-\8}^{+\8} d t'
f'(t') \nonumber\\
&&C'_{\mu}(f(t))\, C'_{\mu}(f(t')) \log | t-t'|
\end{eqnarray}
and upon passing to the inverse function \(\tau(f)\),
\begin{eqnarray}
&&A[\tau] = \frac{1}{2\pi}\,  \int_{-\8}^{+\8}d f \int_{-\8}^{+\8} d f'\nonumber\\
&&
 C'_{\mu}(f)\, C'_{\mu}(f') \log |\tau(f)-\tau(f')|
\end{eqnarray}

To switch to the unit disk, map the upper half–plane via
\begin{smalleq}
\xi_1 + \i \xi_2 = \i \frac{1-\omega}{1+\omega} \;; \omega = r e^{\i
\alpha}\; ; r \le 1.
\end{smalleq}
The real axis goes to \(|\omega|=1\). Changing variables,
\begin{eqnarray}
&&X_{\mu}(r,\alpha) =  \Re \int_{-\pi}^{\pi}  \frac{d \theta}{\pi}
C_{\mu}(\phi(\theta))\nonumber\\
&&
\left(\frac{1}{1- r\exp{\i\alpha- \i\theta}}- \frac{1}{1 +
\exp{-\i\theta}}\right)
\end{eqnarray}
with
\begin{smalleq}
\phi(\theta) = f\left(\tan{\frac{\theta}{2}}\right).
\end{smalleq}
The last term is a pure translation and can be dropped. The area becomes
\begin{smalleq}
A[\phi] = \frac{1}{4 \pi} \int_{-\pi}^{\pi} d \theta \int_{-\pi}^{\pi} d
\theta'
\frac{\left(C_{\mu}(\phi(\theta))-
C_{\mu}(\phi(\theta'))\right)^2}{\left| e^{\i \theta} - e^{\i \theta'}
\right|^2}
\end{smalleq}
or, after integration by parts and inversion of the parametrization,
\begin{eqnarray}
&&A[\theta] = \frac{1}{2 \pi} \int_{-\pi}^{\pi} d \phi \int_{-\pi}^{\pi} d \phi'C'_{\mu}(\phi)\,C'_{\mu}(\phi') \nonumber\\
&&\log \left| \sin \frac{\theta(\phi) -
\theta(\phi')}{2} \right|
\end{eqnarray}

Minimization with respect to the boundary map \(f\) (in the upper half–plane
formulation) gives the nonlinear condition
\begin{smalleq}
0 = \Re \int_{-\8}^{+\8} d t' \frac{C_{\mu}(f(t'))\,
C_{\mu}'(f(t))}{(t-t'+\i 0)^2}
\hypertarget{NL}{}
\label{NL}
\end{smalleq}
equivalent to the boundary diagonality \rf{Diag}. 
In virtue of this condition, the in–surface, inward–pointing unit co–normal
\(\nu_\mu\) at the boundary is aligned with \(\partial_2 X_\mu\). Explicitly,
\begin{smalleq}\hypertarget{insidenormal}{}
\label{insidenormal}
n_{\mu}\left(C(f(t))\right) \propto \Re \int_{-\8}^{+\8} d t'
 \frac{C_{\mu}(f(t'))}{(t-t'+\i 0)^2}
\end{smalleq}

Written in terms of \(\tau(f)\), \rf{NL} becomes
\begin{smalleq}
0 = \Re\int_{-\8}^{+\8} d f \frac{ C'_{\mu}(f) C'_{\mu}(f')}{\tau(f)
-\tau(f') + \i 0}
\end{smalleq}
Introduce the analytic functions
\begin{smalleq}\hypertarget{WeierstrassFunc}{}
\label{WeierstrassFunc}
F_{\mu}(z) =
\int_{-\8}^{+\8}\frac{d f}{\pi} \frac{C'_{\mu}(f)}{\tau(f)-z}
\end{smalleq}
which decay as \(z^{-2}\). Their boundary values satisfy
\begin{smalleq}
\Im F_{\mu}(\tau + \i 0) = C_{\mu}'(f) f'(\tau)
\end{smalleq}
giving the implicit relation
\begin{smalleq}\hypertarget{parametrization}{}
\label{parametrization}
\int d \tau \Im F_{\mu}(\tau +\i0) = C_{\mu}(f)
\end{smalleq}
Thus \(\Im F_\mu(\tau+\mathrm{i}0)\) is tangent to the boundary curve, while
\(\Re F_\mu(\tau+\mathrm{i}0)\) points along the in–surface co–normal
(inward direction):
\begin{smalleq}
\Re F_{\mu}(\tau+\mathrm{i}0)\;\propto\;\nu_{\mu}.
\end{smalleq}
Away from the boundary there is no simple local relation between \(F_\mu\)
and \(\partial_a X_\mu\).

Equation \rf{NL} amounts to the continuity of \(F_\mu^2\) across the real
axis:
\begin{smalleq}
 F_{\mu}^2(t+\i0) =  F_{\mu}^2(t-\i0)
\end{smalleq}
A general analytic solution with the correct fall–off is
\begin{smalleq}
F_{\mu}^2(z) = (1+ \omega)^4\,P(\omega);\; \omega = \frac{\i - z}{\i + z}
\end{smalleq}
where \(P(\omega)\) is a series (e.g., a polynomial) convergent for
\(|\omega|\le 1\). Its coefficients are fixed by an algebraic minimization
problem.

Flat loops are particularly simple: the task reduces to a conformal map of
the loop to the unit circle. For the circle,
\begin{smalleq}
C_1 + \i C_2 = \omega;\; F_1 = \i F_2 = - \frac{(1+ \omega)^2}{2};\; P = 0.
\end{smalleq}
Small deformations of the circle (or any planar loop) can then be handled
systematically by perturbation theory.

\end{document}